\documentclass[12pt,a4paper]{article}
\pdfoutput=1
\usepackage{jheppub}
\usepackage[english]{babel}
\usepackage[utf8]{inputenc}
\usepackage{mathrsfs}
\DeclareSymbolFont{bbold}{U}{bbold}{m}{n}
\DeclareSymbolFontAlphabet{\mathbbold}{bbold}
\hypersetup{unicode}

\newcommand{\sdot}{\!\cdot\!}

\makeatletter
\def\@fpheader{\relax}
\makeatother

\title{On timelike supersymmetric solutions of Abelian gauged 5-dimensional supergravity}
\author{Samuele Chimento}
\affiliation{Instituto de F\'{\i}sica Te\'orica UAM/CSIC\\
C/ Nicol\'as Cabrera, 13--15,  C.U.~Cantoblanco, E-28049 Madrid, Spain}
\emailAdd{samuele.chimento@csic.es}
\preprint{IFT-UAM/CSIC-17-040}
\abstract{
We consider 5-dimensional gauged supergravity coupled to Abelian vector multiplets, and we look for supersymmetric solutions for which the 4-dimensional K\"ahler base
space admits a holomorphic isometry. Taking advantage of this isometry, we are able to find several supersymmetric solutions for the \mbox{ST$[2,n_v+1]$} special geometric
model with arbitrarily many vector multiplets. 
Among these there are three families of solutions with $n_v+2$ independent parameters, which for one of the families can be seen to correspond to $n_v+1$ electric charges
and one angular momentum. These solutions generalize the ones recently found for minimal gauged supergravity in JHEP 1704 (2017) 017 and include in particular the general supersymmetric asymptotically-AdS$_5$ black holes of Gutowski and Reall, analogous black hole solutions
with non-compact horizon, the three near horizon geometries themselves, and the singular static solutions of Behrndt, Chamseddine and Sabra.
}

\begin{document}
\maketitle
\flushbottom

%%%%%%%%%%%%%%%%%%%%%%%%%%%%%%%%%%%%%%%%%%%%%%%%%%%%%%%%%%%%%%%%%%%%%%
%%%%%%%%%%%%%%%%%%%%%%%%%%%%%%%%%%%%%%%%%%%%%%%%%%%%%%%%%%%%%%%%%%%%%%
%%%%%%%%%%%%%%%%%%%%%%%%%%%%%%%%%%%%%%%%%%%%%%%%%%%%%%%%%%%%%%%%%%%%%%
%%%%%%%%%%%%%%%%%%%%%%%%%%%%%%%%%%%%%%%%%%%%%%%%%%%%%%%%%%%%%%%%%%%%%%
\section{Introduction}
%%%%%%%%%%%%%%%%%%%%%%%%%%%%%%%%%%%%%%%%%%%%%%%%%%%%%%%%%%%%%%%%%%%%%%
%%%%%%%%%%%%%%%%%%%%%%%%%%%%%%%%%%%%%%%%%%%%%%%%%%%%%%%%%%%%%%%%%%%%%%
%%%%%%%%%%%%%%%%%%%%%%%%%%%%%%%%%%%%%%%%%%%%%%%%%%%%%%%%%%%%%%%%%%%%%%
%%%%%%%%%%%%%%%%%%%%%%%%%%%%%%%%%%%%%%%%%%%%%%%%%%%%%%%%%%%%%%%%%%%%%%

Exact solutions of supergravity theories have been and continue to be instrumental 
in gaining new insights into string theory and related areas of research. In particular 
asymptotically anti-de Sitter solutions, which occur naturally in gauged supergravity, 
are interesting from the point of view of the AdS/CFT correspondence, since in that context 
they can be viewed as gravitational duals of strongly coupled quantum systems living on 
the AdS boundary.

Symmetry has always been one of the main tools in the search for exact solutions of gravity theories, 
since requiring the invariance of the solution under some symmetry transformation can dramatically simplify
the usually formidable task of solving the equations of motion. 

In the supergravity setting it is natural to look for solutions with some unbroken supersymmetry.
This implies that the bosonic equations of motion are related through the Killing Spinor Identities \cite{Kallosh:1993wx},
reducing the problem of solving them to that of solving just a small subset plus the first order supersymmetry equations.

However, while assuming unbroken supersymmetry makes the problem more tractable, it is usually not enough 
to find explicit solutions, and one  has to make some additional assumptions or to impose a specific ansatz 
in order to solve the equations.\footnote{For a comprehensive review of supersymmetric
  solutions of supergravity theories with many references see,
  \textit{e.g.}~Ref.\cite{Ortin:2015hya}.}

An approach that has proven to be very successful in ungauged 5-dimensional supergravity, with or without vector multiplets, is to assume 
that the 4-dimensional base space, which for that theory has to be hyperKähler, admits one triholomorphic isometry.
In this case the base space has a Gibbons-Hawking metric \cite{Gibbons:1979zt,Gibbons:1987sp}, and it turns out that the solutions can be completely characterized
in terms of a small number of building blocks, namely harmonic functions on 3-dimensional flat space \cite{Gauntlett:2002nw, Gauntlett:2004qy}. The same ansatz 
has also been effective for $\mathcal{N}=1$, $d=5$ supergravity with vector multiplets and non-Abelian gaugings \cite{Meessen:2015enl} ,
but without Fayet-Iliopoulos terms, in which case the base space is again a 4-dimensional hyperKähler space.

Recently \cite{Chimento:2016mmd} a similar ansatz was applied to the case of minimal $d=5$ gauged supergravity, where a U(1) subgroup
of the SU(2) R-symmetry group is gauged by adding a Fayet-Iliopoulos term to the bosonic action. In this case the 
base space is just Kähler, instead of hyperKähler, and the ansatz consists in assuming that it admits a holomorphic isometry.
The metric of the base space can then be written in terms of two functions \cite{Chimento:2016run} in a form that generalizes the Gibbons-Hawking metrics,
and the problem of finding supersymmetric solutions is reduced to that of solving a system of fourth order differential equations 
for these two functions plus a third one.

The aim of this paper is to apply the same ansatz in the case of $\mathcal{N}=1$, $d=5$ supergravity 
with vector multiplets and Abelian Fayet-Iliopoulos gaugings, where a U(1) subgroup
of the SU(2) R-symmetry group is gauged with a linear combination of the vector fields of the 
theory, in which case the base space is again Kähler.

The paper is organized as follows. Section \ref{sec:sugra} consists in a quick review of
the theory and the conditions to impose on the fields in order to obtain
(timelike) supersymmetric solutions. In Section \ref{sec:iso} we  
adapt the supersymmetry equations to the assumption that the 4-dimensional Kähler base 
space of the solution admits a holomorphic isometry, after writing the general form 
for a metric of this kind. In Section \ref{sec:solutions}, after making some additional assumptions,
we find several supersymmetric solutions for the special geometric model \mbox{ST$[2,n_v+1]$} 
with an arbitrary number $n_V$ of vector multiplets. Among these are three general classes of superficially asymptotically-AdS solutions
that can be seen as a generalization in the presence of vector multiplets of solutions found recently
for pure gauged supergravity \cite{Chimento:2016mmd}. They are studied in some detail in Subsection \ref{subsec:case1},
where the conserved charges are computed for one of the families, and it is shown that they include as particular cases black holes with
compact or non-compact horizon, as well as static singular solutions. In Subsection \ref{subsec:case2} we give the explicit expression of the fields
for supersymmetric black holes not included in the solutions of Subsection \ref{subsec:case1}, despite being very similar to a subcase of them.
We conclude in Section \ref{sec-conclusions} with some final remarks.

%%%%%%%%%%%%%%%%%%%%%%%%%%%%%%%%%%%%%%%%%%%%%%%%%%%%%%%%%%%%%%%%%%%%%%
%%%%%%%%%%%%%%%%%%%%%%%%%%%%%%%%%%%%%%%%%%%%%%%%%%%%%%%%%%%%%%%%%%%%%%
%%%%%%%%%%%%%%%%%%%%%%%%%%%%%%%%%%%%%%%%%%%%%%%%%%%%%%%%%%%%%%%%%%%%%%
%%%%%%%%%%%%%%%%%%%%%%%%%%%%%%%%%%%%%%%%%%%%%%%%%%%%%%%%%%%%%%%%%%%%%%
\section{Abelian gauged \texorpdfstring{$\mathcal{N}=1,d=5$}{N=1, d=5} supergravity\label{sec:sugra}}
%%%%%%%%%%%%%%%%%%%%%%%%%%%%%%%%%%%%%%%%%%%%%%%%%%%%%%%%%%%%%%%%%%%%%%
%%%%%%%%%%%%%%%%%%%%%%%%%%%%%%%%%%%%%%%%%%%%%%%%%%%%%%%%%%%%%%%%%%%%%%
%%%%%%%%%%%%%%%%%%%%%%%%%%%%%%%%%%%%%%%%%%%%%%%%%%%%%%%%%%%%%%%%%%%%%%
%%%%%%%%%%%%%%%%%%%%%%%%%%%%%%%%%%%%%%%%%%%%%%%%%%%%%%%%%%%%%%%%%%%%%%

In this section we give a brief description of the bosonic sector of a general
theory of $\mathcal{N}=1,d=5$ supergravity coupled to $n_{v}$ vector
multiplets in which a U$(1)$ subgroup of the SU$(2)$ R-symmetry group has been
gauged by the addition of Fayet-Iliopoulos (FI) terms. The U$(1)$ subgroup
to be gauged and the gauge vector used in the gauging are determined by the
tensor $\mathsf{P}_{I}{}^{r}$, as we are going to explain.\footnote{Although
  its origin is different, it can be understood as a particular example of
  embedding tensor.}  Our conventions are those in
Refs.~\cite{Bellorin:2006yr,Bellorin:2007yp} which are those of
Ref.~\cite{Bergshoeff:2004kh} with minor modifications.

The supergravity multiplet is constituted by the graviton $e_{\mu}^{a},$ the
gravitino $\psi_{\mu}^{i}$ and the graviphoton $A_{\mu}$.  All the spinors are
symplectic Majorana spinors and carry a fundamental SU(2) R-symmetry index.
The $n_{v}$ vector multiplets, labeled by $x=1,....,n_{v}$ consist of a real
vector field $A_{\mu}^{x},$ a real scalar $\phi^{x}$ and a gaugino
$\lambda^{i\, x}$. 

It is convenient to combine the matter vector fields $A_{\mu}^{x}$ with the graviphoton 
$A_{\mu}\equiv A_{\mu}^{0}$ into a vector $(A_{\mu}^{I})= (A^{0}_{\mu},A^{i}_{\mu})$. 
It is also convenient to define a vector of functions of the scalars
$h^{I}(\phi)$. $\mathcal{N}=1,d=5$ supersymmetry requires that these $n_{v}+1$
functions of the $n_{v}$ scalars satisfy a constraint of the form

\begin{equation}
\label{eq:Chhh}
C_{IJK}h^{I}(\phi)h^{J}(\phi)h^{K}(\phi)=1\, ,
\end{equation}

\noindent
where the constant symmetric tensor $C_{IJK}$ completely characterizes the
ungauged theory and the \textit{Special Real geometry} of the scalar
manifold. In particular, the kinetic matrix of the vector fields
$a_{IJ}(\phi)$ and the metric of the scalar manifold $g_{xy}(\phi)$ can be
derived from it as follows: first, we define

\begin{equation}
h_{I}\equiv C_{IJK}h^{J}h^{K}\, ,
\,\,\,\,\,
\Rightarrow
\,\,\,\,\,
h^{I}h_{I}=1\, ,
\end{equation}

\noindent
and 

\begin{equation}
h^{I}_{x} 
\equiv
-\sqrt{3} h^{I}{}_{,x}
\equiv  
-\sqrt{3} \frac{\partial h^{I}}{\partial\phi^{x}}\, ,  
\hspace{1cm}
h_{Ix}
\equiv  
+\sqrt{3}h_{I, x}\, ,
\,\,\,\,\,
\Rightarrow
\,\,\,\,\,
h_{I}h^{I}_{x}
=
h^{I}h_{Ix}
=
0\, .   
\end{equation}

\noindent
Then, $a_{IJ}$ is defined implicitly by the relations

\begin{equation}
h_{I}  = a_{IJ}h^{I}\, ,
\hspace{1cm}
h_{Ix}  = a_{IJ}h^{J}{}_{x}\, .
\end{equation}

\noindent
It can be checked that

\begin{equation}
a_{IJ}
=
-2C_{IJK}h^{K} +3h_{I}h_{J}\, .  
\end{equation}

The metric of the scalar manifold $g_{xy}(\phi)$, which we will use to raise
and lower $x,y$ indices is (proportional to) the pullback of $a_{IJ}$

\begin{equation}
g_{xy}
\equiv
a_{IJ}h^{I}{}_{x}h^{J}{}_{y}
=
-2C_{IJK}h_{x}^{I}h_{y}^{J}h^{K}\, .
\end{equation}

We will use the completeness relation

\begin{equation}
\label{eq:completeness}
h_{I}h^{J}+g^{xy}h_{x\, I}h_{y}{}^{J} = \delta_{I}{}^{J}\, .  
\end{equation}

The FI gauging of any model of $\mathcal{N}=1,d=5$ supergravity coupled to
vector multiplets is completely determined by the choice of
$\mathsf{P}_{I}{}^{r}$, where $r=1,2,3$ is a $\mathfrak{su}(2)$ index. In the
Abelian case, this tensor can be factorized as follows:

\begin{equation}
\mathsf{P}_{I}{}^{r} = g c_{I}d^{r} \equiv g_{I} d^{r}\, ,  
\end{equation}

\noindent
where $g$ is the gauge coupling constant, $d^{r}$ (which we can normalize
$d^{r}d^{r}=1$) chooses a direction in S$^{3}$ or, equivalently, a
$\mathfrak{u}(1)\subset \mathfrak{su}(2)$ to be gauged and $c_{I}$ (also
normalized $c_{I}c_{I}=1$) dictates which linear combination of the vector
fields, $c_{I}A^{I}{}_{\mu}$, acts as gauge field. $g_{I}=g c_{I}$ is a
convenient combination of constants that we will use. We will not
make any specific choices for the time being.

The bosonic action is given in terms of $a_{IJ},g_{xy}$ and $C_{IJK}$ and
$\mathsf{P}_{I}{}^{r}$

\begin{equation}
\begin{split}
 S  
=
{\displaystyle\int} d^{5}x\sqrt{g}\
\biggl\{
R
+{\textstyle\frac{1}{2}}g_{xy}\partial_{\mu}\phi^{x}
\partial^{\mu}\phi^{y}
-V(\phi)
&-{\textstyle\frac{1}{4}} a_{IJ} F^{I\mu\nu}F^{J}{}_{\mu\nu}\\
\\
&+\frac{C_{IJK}\varepsilon^{\mu\nu\rho\sigma\alpha}}{12\sqrt{3}\sqrt{g}}
F^{I}{}_{\mu\nu}F^{J}{}_{\rho\sigma}A^{K}{}_{\alpha}
\biggr\}\, ,
  \end{split}
\end{equation}

\noindent
where the Abelian vector field strengths are
$F^{I}{}_{\mu\nu}=2\partial_{[\mu}A^{I}{}_{\nu]}$ and the scalar potential
$V(\phi)$ is given by

\begin{equation}
V(\phi) 
= 
-\left(4h^{I}h^{J} -2g^{xy}h_{x}^{I}h_{y}^{J}\right)
\mathsf{P}_{I}{}^{r}\mathsf{P}_{J}{}^{r}
=
-4 C^{IJK}h_I \mathsf{P}_{J}{}^{r}\mathsf{P}_{K}{}^{r}
\, .
\end{equation}

The equations of motion for the bosonic fields are

\begin{align}
G_{\mu\nu}
-{\textstyle\frac{1}{2}}a_{IJ}\left(F^{I}{}_{\mu}{}^{\rho} F^{J}{}_{\nu\rho}
-{\textstyle\frac{1}{4}}g_{\mu\nu}F^{I\, \rho\sigma}F^{J}{}_{\rho\sigma}
\right)\phantom{+{\textstyle\frac{1}{2}}g_{\mu\nu}V = {}}\,\,&      
\nonumber\\
&   \\
+{\textstyle\frac{1}{2}}g_{xy}\left(\partial_{\mu}\phi^{x} 
\partial_{\nu}\phi^{y}
-{\textstyle\frac{1}{2}}g_{\mu\nu}
\partial_\rho\phi^{x} \partial^{\rho}\phi^{y}\right)
+{\textstyle\frac{1}{2}}g_{\mu\nu}V
 = {} &
0\, ,
\nonumber\\ 
&  \nonumber \\
\nabla_{\nu}\left(a_{IJ} F^{J\, \nu\mu}\right)
+{\textstyle\frac{1}{4\sqrt{3}}} 
\frac{\varepsilon^{\mu\nu\rho\sigma\alpha}}{\sqrt{g}}
C_{IJK} F^{J}{}_{\nu\rho}F^{k}{}_{\sigma\alpha}
= {} &
0\, ,
\\
& \nonumber \\
\nabla_{\mu}\partial^{\mu}\phi^{x} 
+{\textstyle\frac{1}{4}}g^{xy} \partial_{y}
a_{IJ} F^{I\, \rho\sigma}F^{J}{}_{\rho\sigma}
+g^{xy}\partial_{y}V
= {} &
0\, .
\end{align}

%%%%%%%%%%%%%%%%%%%%%%%%%%%%%%%%%%%%%%%%%%%%%%%%%%%%%%%%%%%%%%%%%%%%%%
%%%%%%%%%%%%%%%%%%%%%%%%%%%%%%%%%%%%%%%%%%%%%%%%%%%%%%%%%%%%%%%%%%%%%%
%%%%%%%%%%%%%%%%%%%%%%%%%%%%%%%%%%%%%%%%%%%%%%%%%%%%%%%%%%%%%%%%%%%%%%
%%%%%%%%%%%%%%%%%%%%%%%%%%%%%%%%%%%%%%%%%%%%%%%%%%%%%%%%%%%%%%%%%%%%%%
%%%%%%%%%%%%%%%%%%%%%%%%%%%%%%%%%%%%%%%%%%%%%%%%%%%%%%%%%%%%%%%%%%%%%%
\subsection{Timelike supersymmetric solutions}
%%%%%%%%%%%%%%%%%%%%%%%%%%%%%%%%%%%%%%%%%%%%%%%%%%%%%%%%%%%%%%%%%%%%%%
%%%%%%%%%%%%%%%%%%%%%%%%%%%%%%%%%%%%%%%%%%%%%%%%%%%%%%%%%%%%%%%%%%%%%%
%%%%%%%%%%%%%%%%%%%%%%%%%%%%%%%%%%%%%%%%%%%%%%%%%%%%%%%%%%%%%%%%%%%%%%
%%%%%%%%%%%%%%%%%%%%%%%%%%%%%%%%%%%%%%%%%%%%%%%%%%%%%%%%%%%%%%%%%%%%%%
%%%%%%%%%%%%%%%%%%%%%%%%%%%%%%%%%%%%%%%%%%%%%%%%%%%%%%%%%%%%%%%%%%%%%%

The general form of the solutions of these theories admitting a timelike
Killing spinor\footnote{A timelike (commuting) spinor $\epsilon^{i}$ is, by
  definition, such that the real vector bilinear constructed from it
  $iV_{\mu}\sim \bar{\epsilon}_{i}\gamma_{\mu}\epsilon^{i}$ is timelike.} was
found in Refs.~\cite{Gauntlett:2003fk,Gutowski:2004yv,Gutowski:2005id}. In
what follows we are going to review it using the notation and results of
Ref.~\cite{Bellorin:2007yp} in which general non-Abelian gaugings were 
considered,\footnote{Even more general gaugings were considered in \cite{Bellorin:2008we} with the inclusion of tensor multiplets.}
but restricting to Abelian FI gaugings.

The building blocks of the timelike supersymmetric solutions are the scalar
function $\hat{f}$, the 4-dimensional spatial metric
$h_{\underline{m}\underline{n}}$,\footnote{$m,n,p=1,\cdots,4$ will be tangent
  space indices and $\underline{m},\underline{n},\underline{p}=1,\cdots,4$
  will be curved indices. We are going to denote with hats all objects that
  naturally live in this 4-dimensional space.} an antiselfdual almost
hypercomplex structure $\hat{\Phi}^{(r)}{}_{mn}$,\footnote{That is: the 2-forms
  $\hat{\Phi}^{(r)}{}_{mn}$ $r,s,t=1,2,3$ satisfy
\begin{eqnarray}
\hat{\Phi}^{(r)\, mn} 
& = &
-\tfrac{1}{2}\varepsilon^{mnpq}\hat{\Phi}^{(r)}{}_{pq}\, , 
\hspace{1cm}
\mbox{or}
\hspace{1cm}
\hat{\Phi}^{(r)}=-\star_{4}\hat{\Phi}^{(r)}\, ,
\\
& & \nonumber \\
\hat{\Phi}^{(r)\, m}{}_{n}  \hat{\Phi}^{(s)\, n}{}_{p}
& = &
-\delta^{rs} \delta^{m}{}_{p} 
+\varepsilon^{rst}  \hat{\Phi}^{(t)\, m}{}_{p}\, .
\end{eqnarray}
} a 1-form $\hat{\omega}_{\underline{m}}$, the 1-form potentials
$\hat{A}^{I}{}_{\underline{m}}$ and the scalars of the theory combined into
the functions $h^{I}(\phi)$. All these fields are defined on the 4-dimensional spatial
manifold usually called ``base space''. They are time-independent and must
satisfy a number of conditions:

\begin{enumerate}
\item The antiselfdual almost hypercomplex structure $\hat{\Phi}^{(r)}{}_{mn}$, the
  1-form potentials $\hat{A}^{I}{}_{\underline{m}}$ and the base space metric
  $h_{\underline{m}\underline{n}}$ (through its Levi-Civita connection)
  satisfy the equation

\begin{equation}
\label{eq:dfAf}
\hat{\nabla}_{m}\hat{\Phi}^{(r)}{}_{np} 
+
\varepsilon^{rst}\hat{A}^{I}{}_{m}\mathsf{P}_{I}{}^{s}\hat{\Phi}^{(t)}{}_{np}
=0\, .
\end{equation}

\item The selfdual part of the spatial vector field strengths
  $\hat{F}^{I}\equiv d\hat{A}^{I}$ must be related to the  function $\hat{f}$,
  the 1-form $\hat{\omega}$ and the scalars of the theory by 

\begin{equation}
\label{eq:susy_{2}}
h_{I}\hat{F}^{I+} 
= 
{\textstyle\frac{2}{\sqrt{3}}} (\hat{f}d\hat{\omega})^{+} \, , 
\end{equation}

\item while the antiselfdual part is related to the almost hypercomplex
  structure by\footnote{In this equation the indices of $C^{IJK}$ have been
    raised using the inverse metric $a^{IJ}$ and one has the useful relations
\begin{equation}
\label{eq:Ch}
C^{IJK}h_{K}
=
h^{I}h^{J} -\tfrac{1}{2}g^{xy}h_{x}^{I}h_{y}^{J}
=\tfrac{3}{2}h^{I}h^{J}-\tfrac{1}{2}a^{IJ}\, .
\end{equation}
}

\begin{equation}
\label{eq:Fminus}
\hat{F}^{I-} 
=
-2\hat{f}^{-1}C^{IJK}h_{J}\mathsf{P}_{K}{}^{r}\hat{\Phi}^{(r)}\, .
\end{equation}

\item Finally, all the building blocks are related by the equation

\begin{equation}
\label{eq:susy_4}
\hat{\nabla}^{2}\left(h_{I}/\hat{f}\right)
-\tfrac{1}{6}C_{IJK}
\hat{F}^{J}\cdot\hat{\star}\hat{F}^{K}
+{\textstyle\frac{1}{2\sqrt{3}}}
\left(a_{IK}-2C_{IJK}h^{J}\right)\hat{F}^{K}\cdot (\hat{f}d\hat{\omega})^{-}
=
0\, ,     
\end{equation}

\noindent
where the dots indicate standard contraction of all the indices of the tensors.

\end{enumerate}

Once the building blocks that satisfy the above conditions have been found,
the physical 5-dimensional fields can be built out of them\footnote{In the
  ungauged case the above conditions determine the quotients $h_{I}/\hat{f}$
  from which $\hat{f}$ can be found by using the condition
  Eq.~(\ref{eq:Chhh}).} as follows: 

\begin{enumerate}
\item The 5-dimensional (conformastationary) metric is given by

\begin{equation}
\label{eq:metric_form}
  ds^{2} 
  = 
  \hat{f}^{\, 2}(dt+\hat{\omega})^{2}
  -\hat{f}^{\, -1}h_{\underline{m}\underline{n}}dx^{m} dx^{n}\, .
\end{equation}

\item The complete 5-dimensional vector fields are given by

\begin{equation}
  \label{eq:completevectorfields}
  A^{I} 
  = 
  -\sqrt{3}h^{I}e^{0} +\hat{A}^{I}\, ,    
  \,\,\,\,\,
  \mbox{where}
  \,\,\,\,\,
  e^{0} 
  \equiv
  \hat{f} (dt +\hat{\omega})\, ,
\end{equation}

\noindent
so that the spatial components are

\begin{equation}
  A^{I}{}_{\underline{m}} 
  = 
  \hat{A}^{I}{}_{\underline{m}} -\sqrt{3}h^{I}\hat{f} \hat{\omega}_{\underline{m}}\, ,
\end{equation}

\noindent
and the 5-dimensional field strength is

\begin{equation}
\label{eq:gauge_field_strenghts}
F^{I}  
= 
-\sqrt{3} d(h^{I} e^{0})  +\hat{F}^{I}\, .
\end{equation}

\item The scalar fields $\phi^{x}$ can be obtained by inverting the functions
  $h_{I}(\phi)$ or $h^{I}(\phi)$. A parametrization which is always available
  is

\begin{equation}
\label{eq:phys_scalars}
  \phi^{x}= h_{x}/h_{0}\, .    
\end{equation}

\end{enumerate}

As it has already been observed in
Refs.~\cite{Gauntlett:2003fk,Gutowski:2005id} choosing
$d^{r}=\delta^{r}{}_{1}$ we see that Eq.~(\ref{eq:dfAf}) gives us additional
information: it splits into

\begin{eqnarray}
\label{eq:df1=0}
\hat{\nabla}_{m}\hat{\Phi}^{(1)}{}_{np} 
& = &
0\, , 
\\
& & \nonumber \\
\label{eq:df2=pf3}
\hat{\nabla}_{m}\hat{\Phi}^{(2)}{}_{np} 
& = & 
\hat{P}_{m}\hat{\Phi}^{(3)}{}_{np}\, ,
\\
& & \nonumber \\
\label{eq:df3=-pf2}
\hat{\nabla}_{m}\hat{\Phi}^{(3)}{}_{np} 
& = & 
-\hat{P}_{m}\hat{\Phi}^{(2)}{}_{np}\, ,
\end{eqnarray}

\noindent
where we have defined

\begin{equation}
\label{eq:Pdef}
\hat{P}_{m}
\equiv
g_{I}\hat{A}^{I}{}_{m}\, .
\end{equation}

The first equation means that the metric $h_{\underline{m}\underline{n}}$ is
K\"ahler with respect to the complex structure
$\hat{J}_{mn}\equiv\hat{\Phi}^{(1)}{}_{mn}$. Taking this fact into
account,\footnote{We use the integrability condition of Eq.~(\ref{eq:df1=0})
\begin{equation}
\hat{R}_{mnpq}=\hat{R}_{mnrs}\hat{J}^{r}{}_{p}\hat{J}^{s}{}_{q}\, ,    
\end{equation}
which leads to the relation between the Ricci and Riemann tensors
\begin{equation}
\hat{R}_{mn}=-\tfrac{1}{2}\hat{R}_{mprq}\hat{J}^{rq}\hat{J}^{p}{}_{n}\, .  
\end{equation}
The Ricci 2-form, defined as
\begin{equation}
\hat{\mathfrak{R}}_{mn}\equiv \hat{R}_{mp}\hat{J}^{p}{}_{n}\, ,  
\end{equation}
is, therefore, related to the Riemann tensor by 
\begin{equation}
\hat{\mathfrak{R}}_{mn} = \tfrac{1}{2}\hat{R}_{mnpq}\hat{J}^{pq}\, .  
\end{equation}
} the integrability condition of the other two equations is\footnote{If
  $P_{m}$ vanishes (for instance, in the ungauged case), then we have a
  covariantly constant hyper-K\"ahler structure and, then, the base space is
  hyperK\"ahler.}

\begin{equation}
\label{eq:R=dP}
\hat{\mathfrak{R}}_{mn} 
= 
-2\hat{\nabla}_{[m}\hat{P}_{n]}
=
-g_{I}\hat{F}^{I}{}_{mn}\, .  
\end{equation}

This equation must be read as a constraint on the 1-form potentials
$\hat{A}^{I}_{\underline{m}}$ posed by the choice of base space metric.

Eq.~(\ref{eq:Fminus}) takes a simpler form as well:

\begin{equation}
\label{eq:susy_{3}}
\hat{F}^{I-}
=
-2\hat{f}^{-1}C^{IJK}h_{J}g_{K}\hat{J}\, ,
\hspace{1cm}
\Rightarrow
\hspace{1cm}
\left\{
  \begin{array}{rcl}
g_{I}\hat{F}^{I-} 
& = & 
\tfrac{1}{2}\hat{f}^{-1}V(\phi) \hat{J}\, ,
\\
& & \\
h_{I}\hat{F}^{I-}
& = &
-2\hat{f}^{-1}g_{I}h^{I}\hat{J}\, .
\end{array}
\right.
\end{equation}

Tracing the first of these equations and Eq.~(\ref{eq:R=dP}) with
$\hat{J}^{mn}$ one finds a relation between the Ricci scalar of the base space
metric, the scalar potential and the function $\hat{f}$:

\begin{equation}
\label{eq:RicciVf}
  \hat{R}= - 2V/\hat{f}\, .
\end{equation}

The last equation to be simplified by our choice is
Eq.~(\ref{eq:susy_4}). Substituting in it Eq.~(\ref{eq:susy_{3}}) and using
Eqs.~(\ref{eq:Ch}) and the completeness relation Eq.~(\ref{eq:completeness})
one finds

\begin{equation}
\label{eq:susy_4-2}
\hat{\nabla}^{2}\left(h_{I}/\hat{f}\right)
-\tfrac{1}{6}C_{IJK}
\hat{F}^{J}\cdot\hat{\star}\hat{F}^{K}
+\tfrac{1}{\sqrt{3}}g_{I}
\hat{J}\cdot (d\hat{\omega})
=
0\, . 
\end{equation}

In order to make progress one has to start making specific assumptions about
the base space metric. In the ungauged \cite{Gauntlett:2002nw,Bellorin:2006yr}
and the non-Abelian gauged cases \cite{Meessen:2015enl} it has proven very
useful to assume that the base space metric has an additional isometry
because, then, it depends on a very small number of independent functions.
Recently the same assumption was made for pure gauged supergravity \cite{Chimento:2016mmd}, 
where the base space can be a general K\"ahler metric, 
allowing to reduce the problem of finding supersymmetric solutions to a system 
of fourth order differential equations for three functions. In what follows we are 
going to make the same assumption for the case at hand, in which vector multiplets are 
present, in the attempt to simplify the task of finding supersymmetric solutions.

%%%%%%%%%%%%%%%%%%%%%%%%%%%%%%%%%%%%%%%%%%%%%%%%%%%%%%%%%%%%%%%%%%%%%%
%%%%%%%%%%%%%%%%%%%%%%%%%%%%%%%%%%%%%%%%%%%%%%%%%%%%%%%%%%%%%%%%%%%%%%
%%%%%%%%%%%%%%%%%%%%%%%%%%%%%%%%%%%%%%%%%%%%%%%%%%%%%%%%%%%%%%%%%%%%%%
%%%%%%%%%%%%%%%%%%%%%%%%%%%%%%%%%%%%%%%%%%%%%%%%%%%%%%%%%%%%%%%%%%%%%%
\section{Timelike supersymmetric solutions of Abelian gauged \texorpdfstring{$\mathcal{N}=1, d=5$}{N=1, d=5}
  supergravity with one additional isometry\label{sec:iso}}
\label{eq:susysolutions}
%%%%%%%%%%%%%%%%%%%%%%%%%%%%%%%%%%%%%%%%%%%%%%%%%%%%%%%%%%%%%%%%%%%%%%
%%%%%%%%%%%%%%%%%%%%%%%%%%%%%%%%%%%%%%%%%%%%%%%%%%%%%%%%%%%%%%%%%%%%%%
%%%%%%%%%%%%%%%%%%%%%%%%%%%%%%%%%%%%%%%%%%%%%%%%%%%%%%%%%%%%%%%%%%%%%%
%%%%%%%%%%%%%%%%%%%%%%%%%%%%%%%%%%%%%%%%%%%%%%%%%%%%%%%%%%%%%%%%%%%%%%
Any four-dimensional K\"ahler metric with one holomorphic isometry can be
written locally as \cite{Chimento:2016run}:
\begin{equation}
\label{eq:final_metric} 
ds^{2} 
= 
H^{-1}\left( dz+\chi \right)^{2}
+H\left\{(dx^{2})^{2}+W^{2}[(dx^{1})^{2}+(dx^{3})^{2}]\right\}\,,
\end{equation} 
with the functions $H$ and $W$, and the 1-form $\chi$, depending only on the
three coordinates $x^{i}$ and satisfying the constraints:
\begin{equation}
\label{eq:constraintijcurved}
\begin{array}{rcl}
(d\chi)_{\underline{1}\underline{2}} 
& = &
\partial_{\underline{3}}H\, ,
\\
& & \\
(d\chi)_{\underline{2}\underline{3}} 
& = &
\partial_{\underline{1}}H\, ,
\\
& & \\
(d\chi)_{\underline{3}\underline{1}} 
& = &
\partial_{\underline{2}}\left(W^{2}H\right)\, ,
\end{array}
\end{equation}
whose integrability condition is
\begin{equation}
\label{eq:integrability}
\mathfrak{D}^{2}H\equiv \partial_{\underline{1}}^2 H
+\partial_{\underline{2}}^2\left(W^{2} H\right)
+\partial_{\underline{3}}^2 H 
= 0\, .
\end{equation}
In a frame defined by the Vierbein
\begin{align}
 e^{\sharp}&=H^{-1/2}\left( dz+\chi \right)\,,\\
 \nonumber\\
 e^{2}&=H^{1/2}dx^2\,,\\
 \nonumber\\
 e^{1,3}&=H^{1/2}W dx^{1,3}\,,
\end{align}
the conserved complex structure is given by
\begin{equation}
\label{eq:J_matrix}
 (\hat J_{mn})=
 \begin{pmatrix}
    \phantom{-}0_{2\times 2} & \phantom{-}\mathbbold{1}_{2\times 2}\\
    -\mathbbold{1}_{2\times 2}  & \phantom{-} 0_{2\times 2}
   \end{pmatrix}\,.
\end{equation}
The Ricci tensor and Ricci scalar of the 4-dimensional metric can be expressed in terms of the functions $H$ and $W^2$ in a compact form,
\begin{equation}
\label{eq:Rmn}
\hat{R}_{mn} 
= 
\hat{\nabla}_{m} \hat{\nabla}_{n}\log{W} 
+\hat{J}_{m}{}^{p}\hat{J}_{n}{}^{q}
\hat{\nabla}_{p} \hat{\nabla}_{q}\log{W}\, ,
\hspace{1cm}
\hat{R} 
= 
\hat{\nabla}^{2}\log{W^{2}}\,,
\end{equation}
where the 4-dimensional Laplacian acts on $z$-independent functions as
\begin{equation}
 \hat{\nabla}^{2}f=H^{-1}\overline{\nabla}^{2}f=\frac{1}{HW^2}\left[\partial_{\underline{1}}^2 f+\partial_{\underline{2}}\left(W^{2}\partial_{\underline{2}} f\right)
 +\partial_{\underline{3}}^2 f\right]\,,
\end{equation}
and $\overline{\nabla}^{2}$ is the Laplacian operator associated with the 3-dimensional metric
\begin{equation}
 ds_3^2=(dx^{2})^{2}+W^2[(dx^{1})^{2}+(dx^{3})^{2}]\,.
\end{equation}
The expression for the Ricci scalar should be compared with
Eq.~(\ref{eq:RicciVf}).

We will take the base space metric $h_{\underline{m}\underline{n}}dx^{m} dx^{n}$ to be of the form (\ref{eq:final_metric}), and we will make the identification
$\hat\Phi^{(1)}=\hat J$. We can solve for $\hat{P}_{m}$ in Eqs.~(\ref{eq:df2=pf3}) and
(\ref{eq:df3=-pf2}) if we choose a particular form for the complex structures
$\hat{\Phi}^{(2,3)}$. Without loss of generality they can be chosen to be 
\begin{equation}
\label{eq:J23_matrices} 
(\hat{\Phi}^{(2)}{}_{mn})
=
\begin{pmatrix} 
i\sigma_{2} & \phantom{-i}0_{2\times 2}\\ 
\phantom{i}0_{2\times 2} & - i\sigma_{2} \\
\end{pmatrix}\, ,
\hspace{1.5cm}
(\hat{\Phi}^{(3)}{}_{mn})
=
\begin{pmatrix} 
\phantom{-i}0_{2\times 2} & -i\sigma_{2} \\ 
- i\sigma_{2} & \phantom{-i}0_{2\times 2} \\
\end{pmatrix}\, ,
\end{equation} 
where $\sigma_{2}$ is the second Pauli matrix
\begin{equation}
 \sigma_2=
 \begin{pmatrix}
  0 & -i\\
  i & \phantom{-}0
 \end{pmatrix}\,.
\end{equation}
Then we find that the flat components of $P$ can be written in the compact form 
\begin{equation}
\label{eq:P_vector} 
\hat{P}_{m} = \hat{J}_{m}{}^{n}\,\partial_{n}\log{W}\, .
\end{equation}
On the other hand, recalling the definition of $\hat{P}_{m}$
Eq.~(\ref{eq:Pdef}) we find for the gauge vector and its field strength

\begin{eqnarray}
g_{I}\hat{A}^{I}{}_{m}
& = &
\hat{J}_{m}{}^{n}\,\partial_{n}\log{W}\, ,
\\
& & \nonumber \\
g_{I}\hat{F}^{I}{}_{mn}\label{eq:GIFI}
& = &
-\mathfrak{R}_{mn}
=
-2\hat{\nabla}_{[m|} \hat{\nabla}_{p}\log{W}\hat{J}{}^{p}{}_{|n]}\, .   
\end{eqnarray}

Every (anti-)selfdual 2-form $\mathcal{F}^{\pm}$ on the four dimensional
K\"ahler base space can be written in terms of a 1-form living on the
3-dimensional space $\vartheta=\vartheta_{\underline{i}}dx^{i}$ as

\begin{equation}
\mathcal{F}^{\pm}= \left( dz+\chi \right)\wedge \vartheta \pm H
\star_{3}\vartheta\, .
\end{equation}
The 2-forms we consider here are also $z$-independent and so will the
components of the corresponding 1-forms be. Thus, we introduce the
$z$-independent 3-dimensional 1-forms $\Lambda^{I}$, $\Sigma^{I}$,
$\Omega_{\pm}$ defined by 

\begin{eqnarray}
\label{eq:simp_F+}
\hat{F}^{I+}
& = &
-\tfrac{1}{2} \left( dz+\chi \right)\wedge \Lambda^{I}
-\tfrac{1}{2} H \star_{3}\Lambda^{I}\, ,
\\
& & \nonumber \\
\label{eq:simp_F-}
\hat{F}^{I-}
& = &
-\tfrac{1}{2} \left( dz+\chi \right)\wedge \Sigma^{I}
+\tfrac{1}{2} H \star_{3}\Sigma^{I}\, ,
\\
& & \nonumber \\
(d\hat{\omega})^{\pm}
& = &
\left(dz+\chi \right)\wedge \Omega^{\pm} \pm H \star_{3}\Omega^{\pm}\, ,
\end{eqnarray}

Comparing the expression of $\hat{F}^{I-}$ with Eq.~(\ref{eq:susy_{3}}) and
those of $h_{I}\hat{F}^{I+}$ and $(d\omega)^{+}$ with Eq.~(\ref{eq:susy_{2}}) we
conclude that

\begin{eqnarray}
\label{eq:simp_Sigma_f}
\Sigma^{I} 
& = & 
4\hat{f}^{-1} C^{IJK}h_{J}\, g_{K}\, dx^{2}\, ,
\\
& & \nonumber \\
\label{eq:Omega+}
\Omega^{+}
& = &
-\tfrac{\sqrt{3}}{4} \hat{f}^{-1} h_{I}\Lambda^{I}\, .
\end{eqnarray}

Requiring the closure of the 2-forms $\hat{F}^{I}=\hat{F}^{I+}+\hat{F}^{I-}$
one gets

\begin{equation}
d\left( \Lambda^{I}+\Sigma^{I} \right)=0\, ,
\end{equation}

\noindent
which means that, locally, 

\begin{equation}
\label{eq:simp_Lambda_dW_Sigma}
\Lambda^{I}=d \left( K^{I}/H\right)-\Sigma^{I}\, ,
\end{equation}

\noindent
for some functions $K^{I}$.

From the same condition, using Eq.~(\ref{eq:integrability}) and the definition
of the operator $\mathfrak{D}^{2}$ in that equation, one also gets

\begin{equation}
\label{eq:KI_laplacian}
\mathfrak{D}^{2} K^{I} 
=
2\, \partial_{\underline{2}}\left( H W^{2}\Sigma^{I}_{\underline{2}}\right)\, .
\end{equation}

Using Eq.~(\ref{eq:GIFI}) and its full contraction with $\hat J$ one finds

\begin{equation}
\label{eq:sigma_k_comb}
2g_{I}\Sigma^{I}_{\underline{2}} =\hat\nabla^{2}\log{W^2} \, ,
\hspace{1.5cm} 
g_{I} K^{I} =  \partial_{\underline{2}}\log{W^2}\, ,
\end{equation}

\noindent
where an integration constant reflecting the possibility of adding to the solutions $K^{I}$ of eq.
(\ref{eq:KI_laplacian}) solutions of the homogeneous equation has been set to zero without loss of generality, since from 
\eqref{eq:simp_Lambda_dW_Sigma} it is clear that the $K^I$'s are defined up to a constant times $H$.
Using these relations, Eq.~(\ref{eq:KI_laplacian}) contracted with $g_{I}$ is
automatically satisfied, leaving $n_{V}$ independent equations.
   
It is convenient to rewrite $\hat{\omega}$ as

\begin{equation}
\label{eq:omega_split}
\hat{\omega} 
= 
\omega_{z} \left( dz+\chi \right)+\omega\, , 
\hspace{1cm}
\omega = \omega_{\underline{i}}dx^{i}\, ,
\end{equation}

\noindent
in terms of which 

\begin{equation}
\label{eq:omegapm}
\Omega^{\pm} 
= \pm\tfrac{1}{2}H^{-1}\left( \omega_{z} \star_{3} d\chi+\star_{3}  d\omega\right)
-\tfrac{1}{2} d\omega_{z}\, .
\end{equation}

\noindent
From Eqs.~(\ref{eq:Omega+}) and (\ref{eq:simp_Lambda_dW_Sigma}) we find that 

\begin{equation}
\Omega^{+} 
= 
-\tfrac{\sqrt{3}}{4}\frac{h_{I}}{\hat{f}}
\left[d\left(K^{I}/H\right)-\Sigma^{I}\right]\, ,  
\end{equation}

\noindent
and, then, from Eq.~(\ref{eq:omegapm}) we find that 

\begin{equation}
\Omega^{-} 
= 
-\Omega^{+} -d\omega_{z}
=
\tfrac{\sqrt{3}}{4}\frac{h_{I}}{\hat{f}}
\left[d\left(K^{I}/H\right)-\Sigma^{I}\right]-d\omega_{z}\, .
\end{equation}

\noindent
Using either of the last two equations in Eq.~(\ref{eq:omegapm}) one gets an
equation for $\omega$:

\begin{equation}
\label{eq:simp_dhatomega}
d\omega
=
H\star_{3}d\omega_{z}-\omega_{z} d\chi 
-\tfrac{\sqrt{3}}{2}\frac{h_{I}}{\hat{f}}H\star_{3}
\left[d\left(K^{I}/H \right)-\Sigma^{I} \right]\, .
\end{equation}

Before calculating its integrability condition it is convenient to make a
change of variables (identical to the one made in the ungauged case) to
(partially) ``symplectic-diagonalize'' the right-hand side. Thus, we define
$L_{I}$ and $M$ through

\begin{equation}
\label{eq:newvariables}
\begin{array}{rcl}
h_{I}/\hat{f}
& \equiv &
L_{I} + \tfrac{1}{12}C_{IJK}K^{J}K^{K}/H\, ,
\\    
& & \\
\omega_{z}
& \equiv & 
M +\tfrac{\sqrt{3}}{4} L_{I}K^{I}/H
+\tfrac{1}{24\sqrt{3}}C_{IJK}K^{I}K^{J}K^{K}/H^{2}\, .
\end{array}
\end{equation}

Substituting these two expressions into Eq.~(\ref{eq:simp_dhatomega}) and
using the relation between the 1-form $\chi$ and the functions $H$ and $W$,
Eqs.~(\ref{eq:constraintijcurved}), the equation for $\omega$ takes the
form\footnote{We have left one $\omega_{z}$ in order to get a more compact
  expression.}

\begin{equation}
\begin{split}
d\omega
=
\star_{3}
\Big\{
HdM -MdH &+\tfrac{\sqrt{3}}{4} \left(K^{I}dL_{I}-L_{I}dK^{I} \right)\\
\\
&-H\left(\omega_{z}\partial_{\underline{2}}\log{W^{2}} 
-2\sqrt{3}h^{I}g_{I} \hat{f}^{-2} \right)dx^{2}\Big\}\, , 
\end{split} 
\end{equation}

\noindent
and its integrability equation is just\footnote{One has $\star_{3}d\star_{3}d =
  \overline{\nabla}^{2}$.}

\begin{equation}
\label{eq:integrabilityomegahat}
  \begin{split}
H\overline{\nabla}^{2}M -M\overline{\nabla}^{2}H 
&+\tfrac{\sqrt{3}}{4} \left(K^{I}\overline{\nabla}^{2}L_{I}
-L_{I}\overline{\nabla}^{2}K^{I} \right)\\
\\
&-{\displaystyle\frac{1}{W^{2}}}
\partial_{\underline{2}}\left\{
HW^{2}\left(\omega_{z}\partial_{\underline{2}}\log{W^{2}} 
-2\sqrt{3}h^{I}g_{I} \hat{f}^{-2} \right)\right\}
 =  0\, .  
\end{split}
\end{equation}

This equation can be simplified by using the equations satisfied by the
functions $H$ and $K^{I}$ (\ref{eq:integrability}) and
(\ref{eq:KI_laplacian}), respectively. We postpone doing this until we derive
the equation for the functions $L_{I}$, which follows from
Eq.~(\ref{eq:susy_4-2}). First of all, observe that, with our choice of
complex structure Eq.~(\ref{eq:J_matrix}) 

\begin{equation}
\hat{J}\cdot (d\hat{\omega}) 
= 
4(d\hat{\omega})^{-}_{02} 
= 
4 \Omega^{-}_{\underline{2}}  
=
\sqrt{3}\frac{h_{I}}{\hat{f}}
\left[\partial_{\underline{2}}\left(K^{I}/H\right)-\Sigma^{I}_{\underline{2}}\right]
-\partial_{\underline{2}}\omega_{z}\, .
\end{equation}

\noindent
On the other hand, we have

\begin{align}
\hat{\nabla}^{2}\left(h_{I}/\hat{f}\right)
& = 
{\displaystyle
\frac{1}{H}
\overline{\nabla}^{2}
\left(h_{I}/\hat{f}\right)\, ,
}
\nonumber\\
& \nonumber\\
\hat{F}^{J}\cdot\hat{\star}\hat{F}^{K}
 & = 
{\displaystyle
\Lambda^{J}_{m}\Lambda^{K}_{m}  
-\Sigma^{J}_{m}\Sigma^{K}_{m}
=
\partial_{m}\frac{K^{J}}{H}  \partial_{m}\frac{K^{K}}{H}
-2 \partial_{m}\frac{K^{(J}}{H}\Sigma^{K)}_{m}\, ,
}
\\
& \nonumber\\
{\displaystyle
C_{IJK}H\partial_{m}\frac{K^{J}}{H}  \partial_{m}\frac{K^{K}}{H}
}
& =  
{\displaystyle C_{IJK}\left[
\overline{\nabla}^{2} \left(\frac{K^{J}K^{K}}{2H}\right)
+\frac{K^{J}K^{K}}{2H^{2}}\overline{\nabla}^{2}H
-\frac{K^{J}\overline{\nabla}^{2}K^{K}}{H}\right]\, ,
}\nonumber
\end{align}

\noindent
and, using all these partial results into Eq.~(\ref{eq:susy_4-2}), and (not
everywhere, for the sake of simplicity) the new variables
Eqs.~(\ref{eq:newvariables}), we arrive at

\begin{equation}
\label{eq:simp_lapl_hI_1}
\begin{array}{rcl}
{\displaystyle
\overline{\nabla}^{2} L_{I} 
-C_{IJK}\left[\tfrac{1}{12}\frac{K^{J}K^{K}}{H^{2}}\overline{\nabla}^{2}H  
+\tfrac{1}{6}\frac{K^{J}\overline{\nabla}^{2}K^{K}}{H}
+\tfrac{1}{3}H \partial_{\underline{2}}\left(K^{J}/H\right)\Sigma^{K}_{\underline{2}}\right]
}\hspace{0.5cm}
& & 
\\ 
& & \\
{\displaystyle
+g_{I}H\left\{\frac{h_{L}}{\hat{f}}\left[\partial_{\underline{2}}
    (K^{L}/H)-\Sigma_{\underline{2}}^{L}  \right]
-\tfrac{4}{\sqrt{3}}\partial_{\underline{2}}\omega_{z}  \right\}
}
& = &
0\, .
\end{array}
\end{equation}

We can now use the relation between the 3-dimensional Laplacian and the
$\mathfrak{D}^{2}$ operator and the equations for the 
functions $H$ and $K^{I}$ (\ref{eq:integrability}) and
(\ref{eq:KI_laplacian})

\begin{equation}
  \begin{array}{rcl}
\overline{\nabla}^{2}H  
& = &
{\displaystyle
\frac{\mathfrak{D}^{2}H}{W^{2}}
-\partial_{\underline{2}}H\frac{\partial_{\underline{2}}W^{2}}{W^{2}}
-H\frac{\partial^{2}_{\underline{2}}W^{2}}{W^{2}}
}
=
{\displaystyle
-\partial_{\underline{2}}H\frac{\partial_{\underline{2}}W^{2}}{W^{2}}
-H\frac{\partial^{2}_{\underline{2}}W^{2}}{W^{2}}
}
\, ,
\\
& & \\
\overline{\nabla}^{2}K^{I}  
& = &
{\displaystyle
\frac{\mathfrak{D}^{2}K^{I}}{W^{2}}
-\partial_{\underline{2}}K^{I}\frac{\partial_{\underline{2}}W^{2}}{W^{2}}
-K^{I}\frac{\partial^{2}_{\underline{2}}W^{2}}{W^{2}}
}
\\
& &\\
& &\hspace{2cm} = 
{\displaystyle
\frac{2}{W^{2}} \partial_{\underline{2}}(HW^{2}\Sigma^{I}_{\underline{2}})
-\partial_{\underline{2}}K^{I}\frac{\partial_{\underline{2}}W^{2}}{W^{2}}
-K^{I}\frac{\partial^{2}_{\underline{2}}W^{2}}{W^{2}}
}
\, ,
\\
\end{array}
\end{equation}

\noindent
and the equation for $L_{I}$ becomes

\begin{equation}
\label{eq:maxwell_LI}
\begin{array}{rcl}
{\displaystyle
\overline{\nabla}^{2}L_{I}
+\frac{C_{IJK}}{3W^{2}}\partial_{\underline{2}}\left( W^{2}
  K^{J}\Sigma_{\underline{2}}^{K}-\tfrac{1}{4} H^{-1}K^{J}
  K^{K} \partial_{\underline{2}}W^{2} \right)
}
\hspace{3.5cm}
& & \\
& & \\
{\displaystyle
+g_{I}H\left\{\frac{h_{L}}{\hat{f}}
\left[\partial_{\underline{2}}(K^{L}/H)-\Sigma_{\underline{2}}^{L}  \right]
-\tfrac{4}{\sqrt{3}}\partial_{\underline{2}}\omega_{z}  \right\}
}
& = & 
0\, .
\end{array}
\end{equation}

This equation, once substituted in Eq.~(\ref{eq:integrabilityomegahat}), gives
\begin{equation}
\label{eq:integrabilityomegahat3}
\begin{split}
 \overline{\nabla}^{2}M = -\frac{C_{IJK}}{48\sqrt{3}W^2}\partial_{\underline{2}}\left(H^{-2}K^{I}K^{J}K^{K}\partial_{\underline{2}}W^2\right)
+\frac{C_{IJK}}{8\sqrt{3}}H^{-1}K^{I}K^{J}\partial_{\underline{2}}\Sigma^{K}_{\underline{2}}\hspace{1.2cm}\\
\\
-\frac{\sqrt{3}}{2}\Sigma^{I}_{\underline{2}} \partial_{\underline{2}}L_I
-\frac{\sqrt{3}}{4}\frac{\partial_{\underline{2}}W^2}{W^2}\Sigma^{I}_{\underline{2}}\left(L_{I} - \tfrac{1}{12}C_{IJK}K^{J}K^{K}/H  \right)\,.
\end{split}
\end{equation}

To summarize, to find a solution one would have to solve equations (\ref{eq:integrability}), (\ref{eq:KI_laplacian}), (\ref{eq:maxwell_LI}) and 
(\ref{eq:integrabilityomegahat3}), with $\frac{h_{I}}{\hat{f}}$ and $\omega_z$ given by (\ref{eq:newvariables}), for the functions $H$, $W^2$, $K^I$, 
$\Sigma_{\underline{2}}^I$, $L_I$ and $M$ while imposing the constraints (\ref{eq:simp_Sigma_f}) and (\ref{eq:sigma_k_comb}). This is still a very difficult problem,
in particular because the constraint (\ref{eq:simp_Sigma_f}) involves the symmetric tensor $C^{IJK}$ with raised indices, which in general is not constant and cannot be 
written in a simple way in terms of, for instance, the functions $\frac{h_{I}}{\hat{f}}$.

To simplify the task one could assume that $C^{IJK}$ is constant, as is the case for several interesting models, in which case (\ref{eq:simp_Sigma_f}) and 
(\ref{eq:newvariables}) allow to write $\Sigma_{\underline{2}}^I$ in terms of $H$, $K^I$ and $L_I$. One could then proceed as follows: first choose two functions
$H$ and $W^2$ solving equation (\ref{eq:integrability}), which amounts to choosing a base space, and subsequently solve the system of second order equations 
given by (\ref{eq:KI_laplacian}), (\ref{eq:maxwell_LI}) and (\ref{eq:integrabilityomegahat3}) for $K^I$, $L_I$ and $M$, subject to the algebraic
constraints (\ref{eq:sigma_k_comb}).

Once all these functions are known, eq. (\ref{eq:newvariables}) gives $\frac{h_{I}}{\hat{f}}$ and $\omega_z$, equations (\ref{eq:constraintijcurved}) and 
(\ref{eq:simp_dhatomega}) can be integrated to give respectively $\chi$ and $\omega$, $\hat \omega$ is given by (\ref{eq:omega_split}) and $\hat f$ can be 
obtained from the functions $\frac{h_{I}}{\hat{f}}$ using the special geometric constraint $C^{IJK}h_I h_J h_K=1$. At this point one has all the ingredients 
to write explicitly the metric (\ref{eq:metric_form}), the scalar fields (\ref{eq:phys_scalars}) and the gauge field strengths (\ref{eq:gauge_field_strenghts}), using
equations (\ref{eq:simp_F+}), (\ref{eq:simp_F-}) and (\ref{eq:simp_Lambda_dW_Sigma}).

\section{Solutions\label{sec:solutions}}
Assume\footnote{In what follows we will rename the coordinate $x^2$ to $\varrho$, both for improved readability and for the natural interpretation as ``radial'' coordinate.}
 for simplicity that $H$ only depends on the $\varrho$ coordinate, $H=H(\varrho)$, and that $W^2$ factorizes as $W^2=\Psi(\varrho)\Phi(x^1,x^3)$. 
 Then from (\ref{eq:integrability})
\begin{equation}
 H=\frac{a \varrho + b}{\Psi}\,.
\end{equation}
We will also assume $a\neq 0$, in which case one can set $a=1$ and $b=0$ by shifting and rescaling the coordinate $\varrho$, so that 
\begin{equation}
 H=\frac{\varrho}{\Psi}\,.
\end{equation}
Inspired by the pure supergravity case \cite{Chimento:2016mmd} we will take $\Psi$ to be a third order polynomial in $\varrho$.
In particular eq. (\ref{eq:sigma_k_comb}), which implies
\begin{equation}
 g_I K^I = \frac{\partial_\varrho\Psi}{\Psi}\,,
\end{equation}
suggests to introduce $n_v+1$ polynomials
\begin{equation}
 \Psi^I\equiv\sum_{n=0}^{3}c_n{}^I\varrho^n
\end{equation}
such that $\Psi=g_I\Psi^I$ and
\begin{equation}
 K^I=\frac{\partial_\varrho \Psi^I}{\Psi}\,.
\end{equation}
Eq. (\ref{eq:KI_laplacian}) can be integrated to give
\begin{equation}
\label{eq:sigma_I_poly}
 \Sigma^I_{\underline{2}}=\frac{1}{2\varrho}\left( -\alpha^I+\partial^2_\varrho\Psi^I \right)\,,
\end{equation}
where $\alpha^I$ are integration constants, which we will take to be independent of $x^1$ and $x^3$.
Eq. (\ref{eq:sigma_k_comb}) implies then that $\Phi$ must be a solution of Liouville's equation
\begin{equation}
 \left( \partial_{\underline 1}^2+\partial_{\underline 3}^2 \right)\log\Phi=-2 k\Phi\,,
\end{equation}
with $k$ given by
\begin{equation}
 2 k=g_I\alpha^I\,.
\end{equation}
It is possible to choose without loss of generality $k=0,\pm 1$ and
\begin{equation}
 \Phi=\Phi_{(k)}\equiv\frac{4}{\left\{1+k\left[(x^1)^2+(x^3)^2\right]  \right\}^2}\,.
\end{equation}
Equation (\ref{eq:constraintijcurved}) then determines $\chi$ up to a closed 1-form,
\begin{equation}
 d\chi=\Phi\, dx^3\wedge dx^1 \qquad\Longrightarrow\qquad \chi=\chi_{(k)}\equiv\frac{2\left( x^3 dx^1-x^1 dx^3 \right)}{1+k\left[ (x^1)^2+(x^3)^2 \right]}\,.
\end{equation}

We now focus our attention on special geometric models for which the totally symmetric tensor with raised indices $C^{IJK}$ is 
constant.\footnote{This is the case for instance when the scalar manifold is a symmetric space.}
Comparing the expression for $\Sigma^I$ in (\ref{eq:sigma_I_poly}) with the one in (\ref{eq:simp_Sigma_f}) it seems a natural choice to introduce
$n_v+1$ first order polynomials in $\varrho$, $Q_I$, such that
\begin{equation}
\label{eq:hI_poly_ansatz}
 \frac{h_I}{\hat f}=\frac{Q_I}{8\varrho}\,,\qquad Q_I\equiv q_{0I}+q_{1I}\varrho\,,
\end{equation}
with eq. \eqref{eq:sigma_I_poly} implying the constraints
\begin{align}
 c_3{}^I&=\frac16 C^{IJK}g_J q_{1K}\nonumber\\
 &\label{eq:c2c3_constraints}\\
 c_2{}^I&=\frac12\left( \alpha^I+ C^{IJK}g_J q_{0K}\right)\nonumber\,.
\end{align}
One can then, after computing the functions $L_I$ from the definition (\ref{eq:newvariables}), use equation (\ref{eq:maxwell_LI}) to obtain an
expression for $\partial_\varrho M$. Since the expression must be the same for each of the $n_v+1$ equations (one for each value of $I$), the following 
proportionality conditions must be met:
\begin{align}
 C_{IJK}c_3{}^Jc_3{}^K&\propto g_I\nonumber\\
 \nonumber\\
 4 C_{IJK}c_3{}^J (\alpha^K-2 c_2{}^K)+g_Jc_3{}^Jq_{0I}&\propto g_I\nonumber\\
 \label{eq:prop_conditions}\\
 4 C_{IJK}(\alpha^J-2 c_2{}^J)c_1{}^K+3 g_Jc_1{}^Jq_{0I}&\propto g_I\nonumber\\
 \nonumber\\
 2 C_{IJK}c_1{}^Jc_1{}^K-3 g_Jc_0{}^Jq_{0I}&\propto g_I\nonumber\,.
\end{align}
After this, all that remains to do is to substitute $\partial_\varrho M$ in eq. (\ref{eq:integrabilityomegahat3}) (we also assume for simplicity $M=M(\varrho)$) 
and solve the resulting algebraic equation.

In order to find explicit solutions we will consider a specific model, namely the \mbox{ST$[2,n_v+1]$} model defined by
\begin{equation}
\label{eq:model_def}
 C_{0xy}=C^{0xy}=\frac{\sqrt{3}}{2}\eta_{xy}\,,
\end{equation}
where $x,y=1,\dots,n_v$, $\eta_{xy}$ is the Minkowski $n_v$-dimensional metric, and the other components of $C_{IJK}$ vanish.
This model reduces to pure supergravity for $n_v=1$ and $h^1=h^0$, and includes as a special case the STU model for $n_v=2$.
% , as can be seen defining $\tilde h^1=h^1+h^2$ and $\tilde h^2=h^1-h^2$, so that $C_{IJK}h^I h^J h^K=1$ implies $h^0 \tilde h^1 \tilde h^2=2/(3\sqrt{3})$.
In what follows $x$-type indices will be raised and lowered with $\eta_{xy}$ and their contraction will be denoted by a dot (e.g. $g\sdot c_1\equiv g_x c_1{}^x$).
The constraints (\ref{eq:c2c3_constraints}) become
\begin{equation}
\begin{array}{lcl}
 c_3{}^0=\frac{1}{4\sqrt{3}}g\sdot q_1 &\quad &c_3{}^x=\frac{1}{4\sqrt{3}}\left( g^xq_{10}+g_0q_1{}^x \right)\\
 \\
 c_2{}^0=\frac{\alpha^0}{2}+\frac{\sqrt{3}}{4}g\sdot q_0 & &c_2{}^x=\frac{\alpha^x}{2}+\frac{\sqrt{3}}{4}\left( g^xq_{00}+g_0q_0{}^x \right)\,.
\end{array} 
\end{equation}

The conditions (\ref{eq:prop_conditions}) and equation (\ref{eq:integrabilityomegahat3}) are satisfied for an arbitrary choice of gauging constants $g_I$ only if one of the 
following sets of conditions is met:
\begin{itemize}
 \item[]
 \begin{enumerate}
 \setlength\itemsep{1em}
 \item $q_{00}=\frac{\sqrt{3}}{4}\frac{g_Ic_0{}^I}{(c_1{}^0)^2}\,q_0\sdot q_0\,,\  c_1{}^x=\frac{\sqrt{3}}{2}\frac{g_Ic_0{}^I}{c_1{}^0}\,q_0{}^x\,,\  q_{1x}=\frac{q_{10}}{g_0}\,g_x$      
 \item $q_{1x}=\frac{q_{10}}{g_0}\,g_x\,,\ c_1{}^0=g\sdot q_0=g\sdot c_1=q_0\sdot c_1=c_1\sdot c_1=g_Ic_0{}^I=0$
 \item $q_{1x}=-\frac{q_{10}}{g_0}\,g_x\,,\ q_{0I}=0\ \forall I\,,\ c_1{}^x=0\ \forall x$
 \item $q_{1x}=-\frac{q_{10}}{g_0}\,g_x\,,\ q_{0I}=0\ \forall I\,,\ c_1{}^0=c_1\sdot c_1=0$
 \item $q_1\sdot q_1=-(\frac{q_{10}}{g_0})^2\, g\sdot g\,,\ g\sdot q_1=0\,,\ q_{0I}=0\ \forall I\,,\ c_1{}^x=0\ \forall x$
 \item $q_1\sdot q_1=-(\frac{q_{10}}{g_0})^2\, g\sdot g\,,\ g\sdot q_1=0\,,\ q_{0I}=0\ \forall I\,,\ c_1{}^0=c_1\sdot c_1=0$
 \item $q_{00}=\frac{\sqrt{3}}{4}\frac{g_Ic_0{}^I}{(c_1{}^0)^2}q_0\sdot q_0\,, c_1{}^x=\frac{\sqrt{3}}{2}\frac{g_Ic_0{}^I}{c_1{}^0}q_0{}^x\,, q_{10}=g\sdot q_0=g\sdot q_1=q_0\sdot q_1=q_1\sdot q_1=0$
 \item $q_{10}=c_1{}^0=g\sdot q_0=g\sdot q_1=q_0\sdot q_1=q_1\sdot q_1=g\sdot c_1=q_0\sdot c_1=c_1\sdot c_1=g_Ic_0{}^I=0$
\end{enumerate}
\end{itemize}
For special choices of the gauging there are some other possibilities. 
\begin{itemize}
 \item[]If $g\sdot g=0$:
 \begin{enumerate}
 \setlength\itemsep{1em}
 \item $q_{00}=\frac{\sqrt{3}}{4}\frac{g_Ic_0{}^I}{(c_1{}^0)^2}\,q_0\sdot q_0\,,\ c_1{}^x=\frac{\sqrt{3}}{2}\frac{g_Ic_0{}^I}{c_1{}^0}\,q_0{}^x\,,\  q_{1x}= \beta g_x$
 \item $q_{00}=\frac{\sqrt{3}}{4}\frac{g_Ic_0{}^I}{(c_1{}^0)^2}\,q_0\sdot q_0\,,\ c_1{}^x=\frac{\sqrt{3}}{2}\frac{g_Ic_0{}^I}{c_1{}^0}\,q_0{}^x\,,\  g\sdot q_0=g\sdot q_1=q_0\sdot q_1=q_1\sdot q_1=0$
 \item $c_1{}^0=g\sdot q_0=g\sdot q_1=q_0\sdot q_1=q_1\sdot q_1=g\sdot c_1=q_0\sdot c_1=c_1\sdot c_1=g_Ic_0{}^I=0$
%  \item $c_1\sdot c_1=\sqrt{3}q_{00}g_Ic_0{}^I\,,c_1{}^0=g\sdot q_1=q_1\sdot q_1=0\,,q_{0x}=0\,\forall x\quad \Rightarrow \hat f^{-1}=0\Rightarrow \mbox{NO!!}$
\end{enumerate}
 \item[]If $g_0=0$:
\begin{enumerate}
 \setlength\itemsep{1em}
 \item $q_{00}=\frac{\sqrt{3}}{4}\frac{g \cdot c_0}{(c_1{}^0)^2}\,q_0\sdot q_0\,,\  c_1{}^x=\frac{\sqrt{3}}{2}\frac{g\cdot c_0}{c_1{}^0}\,q_0{}^x\,,\ q_{10}=0$
 \item $q_{10}=c_1{}^0=g\sdot q_0=g\sdot c_1=q_0\sdot c_1=c_1\sdot c_1=g\sdot c_0=0$
\end{enumerate}
 \item[]If $g_0=g\sdot g=0$:
\begin{enumerate}
 \setlength\itemsep{1em}
 \item $q_{00}=\frac{\sqrt{3}}{4}\frac{g \cdot c_0}{(c_1{}^0)^2}\,q_0\sdot q_0\,,\  c_1{}^x=\frac{\sqrt{3}}{2}\frac{g\cdot c_0}{c_1{}^0}\,q_0{}^x$
 \item $c_1{}^0=g\sdot q_0=g\sdot c_1=q_0\sdot c_1=c_1\sdot c_1=g\sdot c_0=0$
\end{enumerate}
 \item[]If $g_x=0\ \forall x$:
\begin{enumerate}
 \setlength\itemsep{1em}
 \item $q_{00}=\frac{\sqrt{3}}{4}\frac{g_0 c_0{}^0}{(c_1{}^0)^2}\,q_0\sdot q_0\,,\  c_1{}^x=\frac{\sqrt{3}}{2}\frac{g_0 c_0{}^0}{c_1{}^0}\,q_0{}^x$
 \item $c_1{}^0=q_0\sdot c_1=c_1\sdot c_1=c_0{}^0=0$
\end{enumerate}
\end{itemize} 

The function $\hat f$ can be computed from (\ref{eq:hI_poly_ansatz}) using the special geometric constraint (\ref{eq:Chhh}), giving
\begin{equation}
 \hat f^{-1}=\frac{\sqrt[3]{C^{IJK}Q_I Q_J Q_K}}{8\varrho}=\frac{\sqrt{3}}{8\varrho}\left[\frac12 \left(q_{00}+q_{10}\varrho\right)\left(q_0\sdot q_0+2 q_0\sdot q_1 \varrho + q_1\sdot q_1 \varrho^2  \right)  \right]^{1/3}\,.
\end{equation}
We are interested in particular in asymptotically anti-de Sitter solutions. Given that the line element of AdS$_5$ (with radius $\ell$) can be written in standard
supersymmetric form as \cite{Chimento:2016mmd}
\begin{equation}
 \begin{split}
  ds^2=\left[dt+\frac{2}{\ell}\varrho\left( dz+\chi_{(k)}\right)\right]^2-&\varrho\left( k+\frac{4}{\ell^2}\varrho\right)\left( dz+\chi_{(k)}\right)^2\\
  \\
 -&\frac{d\varrho^2}{\varrho\left( k+\frac{4}{\ell^2}\varrho\right)} -\varrho\,\Phi_{(k)}\left[(dx^1)^2+(dx^3)^2  \right]\,,
 \end{split}
\end{equation}
one expects that for such solutions as $\varrho\to\infty$ $\hat f$ tends to a constant and $\Psi$ diverges
like $\varrho^3$. These conditions translate to
\begin{equation}
 q_{10}\, q_1\sdot q_1\neq 0\qquad\mbox{and}\qquad g_I c_3{}^I=\frac{1}{4\sqrt{3}}\left( 2 g_0 g\sdot q_1+q_{10}g\sdot g \right)\neq 0\,,
\end{equation}
excluding all the solutions above except the first six for arbitrary gauging. Out of these, however, only the first two are actually asymptotically AdS, 
at least locally, since in the other cases $\omega_z$ does not present the correct behavior, being proportional to $\varrho^{-1}$ (one can also check that 
their scalar curvature does not tend to a constant as $\varrho\to \infty$). In the following we will analyze some properties of these two cases.

\subsection{Case 1\label{subsec:case1}}
We will now analyze in detail the solutions with parameters satisfying the conditions
\begin{equation}
 q_{00}=\frac{\sqrt{3}}{4}\frac{g_Ic_0{}^I}{(c_1{}^0)^2}\,q_0\sdot q_0\,,\qquad  c_1{}^x=\frac{\sqrt{3}}{2}\frac{g_Ic_0{}^I}{c_1{}^0}\,q_0{}^x\,,
 \qquad  q_{1x}=\frac{q_{10}}{g_0}\,g_x\,.
\end{equation}

The functions $\hat f$ and $\Psi$ become
\begin{multline}
\label{eq:c1_f}
 \hat f^{-1}=\frac{\sqrt{3}}{8\varrho}\left[\frac12 \left(\frac{\sqrt{3}}{4}\frac{g_Ic_0{}^I}{(c_1{}^0)^2}q_0\sdot q_0+q_{10}\varrho\right)
 \left(q_0\sdot q_0+2 \frac{q_{10}}{g_0}g\sdot q_0\, \varrho +  \left(\frac{q_{10}}{g_0}\right)^2\! g\sdot g\, \varrho^2  \right)  \right]^{1/3},\\
{} 
\end{multline}
\begin{equation}
\label{eq:c1_psi}
\begin{split}
 \Psi&=\frac{\sqrt{3}}{4}q_{10}g\sdot g\, \varrho^3
 +\left[ k+\frac{\sqrt{3}}{4}\left( 2g_0 g\sdot q_0 +\frac{\sqrt{3}}{4}\frac{g_Ic_0{}^I}{(c_1{}^0)^2}q_0\sdot q_0g\sdot g\right) \right]\varrho^2\\
 \\
 &\hspace{5.7cm}+\left( g_0c_1{}^0+\frac{\sqrt{3}}{2}\frac{g_Ic_0{}^I}{c_1{}^0}g\sdot q_0 \right)\varrho+g_Ic_0{}^I\,,
 \end{split}
\end{equation}
while $\omega_z$ can be obtained from eq. (\ref{eq:newvariables}) after integrating $\partial_\varrho M$,
\begin{equation}
\begin{split}
 \omega_z&=\frac{3}{64}\left( q_0\sdot q_0\frac{g_Ic_0{}^I}{c_1{}^0}\frac{1}{\varrho^2}+\frac{(q_{10})^2}{g_0} g\sdot g\, \varrho \right)+d\\
 &\\
 &\hspace{3.2cm}+\left(q_0\sdot q_0+\frac{4}{\sqrt{3}}\frac{q_{10}}{g_0}c_1{}^0 \right)\left(2g_0+\sqrt{3}\frac{g_Ic_0{}^I}{(c_1{}^0)^2}g\sdot q_0  \right)\frac{3}{256\varrho}
\end{split}
\end{equation}
where $d$ is an arbitrary constant, and $\omega$ from eq. (\ref{eq:simp_dhatomega})
\begin{equation}
 \omega=\left[ \frac{3}{64}\frac{q_{10}}{g_0}\left( 2 g_0 g\sdot q_0+\frac{\sqrt{3}}{4}\frac{g_Ic_0{}^I}{(c_1{}^0)^2}q_0\sdot q_0g\sdot g \right) -d\right]\chi_{(k)}\,.
\end{equation}
Since $\omega$ is of the form $\tilde\omega\chi$ with $\tilde\omega$ constant, it is always possible to reabsorb $\omega$ in $\omega_z$ with a shift in the $t$ coordinate, 
$t\to t+\tilde\omega z$, leading to $\omega=0$ and
\begin{equation}
\label{eq:c1_omz}
 \begin{split}
  \omega_z&=\frac{3}{64 \varrho^2}\Bigg[\frac{(q_{10})^2}{g_0} g\sdot g\, \varrho^3  
 +\frac{ q_{10}}{ g_0}\left( 2 g_0 g\sdot q_0 +\frac{\sqrt{3}}{4}\frac{g_Ic_0{}^I}{(c_1{}^0)^2}q_0\sdot q_0g\sdot g \right)\varrho^2\\
 \\
 &\hspace{1.7cm}+\left(q_0\sdot q_0+\frac{4}{\sqrt{3}}\frac{q_{10}}{g_0}c_1{}^0 \right)\left(2g_0+\sqrt{3}\frac{g_Ic_0{}^I}{(c_1{}^0)^2}g\sdot q_0  \right)\frac{\varrho}{4}
 +q_0\sdot q_0\frac{g_Ic_0{}^I}{c_1{}^0}\Bigg]\,. 
 \end{split}
\end{equation}

The full solution is invariant under the rescaling $t\to t/\alpha$, $\varrho\to \alpha \varrho$, $q_{10}\to q_{10}/\alpha$, $c_1{}^I\to\alpha c_1{}^I$, 
$c_0{}^I\to\alpha^2 c_0{}^I$. Since we are assuming $q_{10}\neq 0$ we can use this freedom to set 
\begin{equation}
\label{eq:q10_rescal}
 q_{10}=\frac{8}{\sqrt{3}}g_0 \ell\,,
\end{equation}
where we introduced for convenience the constant $\ell$ defined by\footnote{The solutions presented here are superficially asymptotically AdS$_5$, with AdS radius $|\ell|$.}
\begin{equation}
 \ell^3 g_0\, g\sdot g=2\,,
\end{equation}
so that $\hat f\to 1$ for $\varrho\to\infty$.

The line element is then
\begin{equation}
\begin{split}
 ds^2&=\hat f^2\left[dt+\omega_z\left( dz+\chi_{(k)}\right)\right]^2\\
 &\\
 &\hspace{1.6cm}-\hat f^{-1}\left\{\frac{\Psi}{\varrho}\left( dz+\chi_{(k)}\right)^2
 +\frac{\varrho}{\Psi}d\varrho^2+\varrho\,\Phi_{(k)}\left[(dx^1)^2+(dx^3)^2\right]\right\}\,,
\end{split}
\end{equation}
with
\begin{align}
 \hat f^{-3}={}&\left(1+\frac{3}{32 g_0 \ell}\frac{g_Ic_0{}^I}{(c_1{}^0)^2}\frac{q_0\sdot q_0}{\varrho}\right)
 \left(1+\frac{\sqrt{3}}{4 g\sdot g \ell}\frac{g\sdot q_0}{\varrho} +  \frac{3}{64 g\sdot g \ell^2} \frac{q_0\sdot q_0}{\varrho^2} \right)\,,\\
 \nonumber\\
 \Psi={}&\frac{4}{\ell^2} \varrho^3
 +\left[ k+\frac{\sqrt{3}}{4}\left( 2g_0 g\sdot q_0 +\frac{\sqrt{3}}{4}\frac{g_Ic_0{}^I}{(c_1{}^0)^2}q_0\sdot q_0g\sdot g\right) \right]\varrho^2\nonumber\\
 \\
 &\hspace{4.6cm}+\left( g_0c_1{}^0+\frac{\sqrt{3}}{2}\frac{g_Ic_0{}^I}{c_1{}^0}g\sdot q_0 \right)\varrho+g_Ic_0{}^I\,,\nonumber\\
 \nonumber\\
 \omega_z={}&\frac{2}{\ell}\varrho
 +\frac{3}{64 \varrho^2}\Bigg[\frac{8\ell}{\sqrt{3}}\left(2 g_0 g\sdot q_0 +\frac{\sqrt{3}}{4}\frac{g_Ic_0{}^I}{(c_1{}^0)^2}q_0\sdot q_0g\sdot g\right)\varrho^2\nonumber\\
 \\
 &\hspace{0.5cm}+\left(q_0\sdot q_0+\frac{32\ell}{3}c_1{}^0 \right)\left(2g_0+\sqrt{3}\frac{g_Ic_0{}^I}{(c_1{}^0)^2}g\sdot q_0  \right)\frac{\varrho}{4}
 +q_0\sdot q_0\frac{g_Ic_0{}^I}{c_1{}^0}\Bigg]\nonumber\,. 
\end{align}
Using the parametrization (\ref{eq:phys_scalars}) the physical scalars are given by
\begin{equation}
 \phi^x=\frac{h_x}{h_0}=\frac{h_x/\hat f}{h_0/\hat f}=\frac{8 g_x \ell \varrho+\sqrt{3} q_{0x}}{8 g_0\ell\varrho+\frac{3}{4}\frac{g_Ic_0{}^I}{(c_1{}^0)^2}q_0\sdot q_0}\,.
\end{equation}
The full gauge potentials are given, according to eq. (\ref{eq:completevectorfields}), by
\begin{equation}
\label{eq:sol_full_gauge_pot}
 A^I=-\sqrt{3}h^I \hat f \left[ dt+\omega_z\left( dz+\chi_{(k)} \right) \right]+\hat A^I\,,
\end{equation}
where the 4-dimensional part $\hat A^I$ can be obtained from (\ref{eq:simp_F+}), (\ref{eq:simp_F-}), (\ref{eq:simp_Lambda_dW_Sigma}),
\begin{align}
 \hat A^0&=\left(   g\sdot g\ell \varrho+\frac{\sqrt{3}}{4}g\sdot q_0 +\frac{1}{2}\frac{c_1{}^0}{\varrho}\right)\left( dz+\chi_{(k)} \right)\,,\\
 \nonumber&\\
 \hat A^x&=\left(2 g_0 g^x\ell\varrho+\frac{3}{16}\frac{g_Ic_0{}^I}{(c_1{}^0)^2}q_0\sdot q_0 g^x+\frac{\sqrt{3}}{4} g_0q_0{}^x
 +\frac{\sqrt{3}}{4} \frac{g_Ic_0{}^I}{c_1{}^0}\frac{q_0{}^x}{\varrho} \right)\left( dz+\chi_{(k)} \right)\,,
\end{align}
while since $h^I=C^{IJK}h_J h_K$\footnote{Note that here $h^x\neq \eta^{xy}h_y$.}
\begin{equation}
 h^0 \hat f = \frac{8 \varrho}{\sqrt{3} (8 g_0\,\ell \varrho+\frac{3}{4}\frac{g_Ic_0{}^I}{(c_1{}^0)^2}q_0\sdot q_0)}\,,\qquad 
 h^x \hat f = \frac{16\varrho}{\sqrt{3}} \frac{8 g^x \ell \varrho+\sqrt{3} q_0{}^x}{(8 g_y \ell \varrho+\sqrt{3} q_{0y})^2}\,.
\end{equation}

Pure supergravity is recovered by choosing $g_x=g_0 \delta^1_x$, $q_{0x}=q_{00}\delta^1_x$ and $q_{1x}=q_{10}\delta^1_x$. With this choice one recovers the class
of asymptotically AdS solutions of minimal gauged $\mathcal{N}=1$, $d=5$ supergravity found in \cite{Chimento:2016mmd}.

For each value of $k$ the solutions are determined by $n_v+2$ parameters, $q_{0x}$, $c_1{}^0$ and $g_Ic_0{}^I$. The metric however only depends on the $q_{0x}$'s
through the combinations $g\sdot q_0$ and $q_0\sdot q_0$, so it is always determined by four parameters, independently of the number of vector multiplets $n_v$.

\subsubsection{Supersymmetric black holes}
If an event horizon exists, it must be situated in $\varrho=0$, where $\hat f=0$ and the supersymmetric Killing vector $\partial_t$ becomes null.
Since $\hat f$, $H$ and $\omega_z$ only depend on $\varrho$, it is possible to perform a coordinate change such that
\begin{align}
 dt&=du-H \hat f^{-1}(\hat f^{-1}H^{-1}-\hat f^2\omega_z{}^2)^{1/2}d\varrho\,,\\
 \nonumber\\
 dz&=dv-\frac{\hat f H \omega_z}{(\hat f^{-1}H^{-1}-\hat f^2\omega_z{}^2)^{1/2}}d\varrho\,,
\end{align}
after which the metric takes the form
\begin{equation}
\begin{split}
\label{eq:null_coord_metric}
 ds^2&=\hat f^2 du^2-\frac{2 du d\varrho}{(\hat f^{-1}H^{-1}-\hat f^2\omega_z{}^2)^{1/2}}+2 \hat f^2 \omega_z du(dv+\chi_{(k)})\\
 \\
 &\hspace{4cm}-(\hat f^{-1}H^{-1}-\hat f^2\omega_z{}^2)(dv+\chi_{(k)})^2-\frac{\varrho}{\hat f}d\Omega^2_{(2,k)}\,.
\end{split}
\end{equation}
The combination $(\hat f^{-1}H^{-1}-\hat f^2\omega_z{}^2)$ tends to a constant in the limit $\varrho\to 0$, so the hypersurface $\varrho=0$
is null, and is thus a Killing horizon, if $\hat f^2 \omega_z$ goes to zero. The only possibility to satisfy 
this condition without giving rise to singularities is to take the scaling limit
\begin{equation}
\label{eq:scaling_limit}
 g_Ic_0{}^I =\frac{4}{\sqrt{3}}\frac{q_{00}}{q_0\sdot q_0}(c_1{}^0)^2\,,\qquad c_1{}^0\to 0\,,
\end{equation}
in which case the functions that determine the metric become
\begin{align}
\label{eq:bh_f}
 \hat f^{-3}&=\left(1+\frac{\sqrt{3}}{8 g_0 \ell}\frac{q_{00}}{\varrho}\right)
 \left(1+\frac{\sqrt{3}}{4 g\sdot g \ell}\frac{g\sdot q_0}{\varrho} +  \frac{3}{64 g\sdot g \ell^2} \frac{q_0\sdot q_0}{\varrho^2} \right)\,,\\
 \nonumber\\
\label{eq:bh_psi}
 \Psi&=\varrho^2\left[\frac{4}{\ell^2} \varrho+ k+\frac{\sqrt{3}}{4}\left( 2g_0 g\sdot q_0 +q_{00} g\sdot g\right) \right]\,,\\
 \nonumber\\
\label{eq:bh_omz}
 \omega_z&=\frac{2}{\ell}\varrho
 +\frac{3}{64 \varrho}\left[\frac{8\ell}{\sqrt{3}}\left( 2 g_0 g\sdot q_0 +q_{00} g\sdot g \right)\varrho+\frac12\left(g_0 q_0\sdot q_0+2 q_{00}g\sdot q_0\right)\right]\,. 
\end{align}
For $k=1$ these are the supersymmetric black holes of \cite{Gutowski:2004yv} with the choice (\ref{eq:model_def}), while for $k=0$ and $k=-1$ one gets a generalization
of the black holes with non-compact horizon found in \cite{Chimento:2016mmd} for pure gauged supergravity.

For them to be regular, any curvature singularity should lie behind the horizon $\varrho=0$. Since the curvature scalars diverge when $\hat f^{-3}$ vanishes, then the zeroes
of (\ref{eq:bh_f}) must be negative, which translates to the conditions
\begin{equation}
 q_{00}\,g\sdot g > 0\,,\qquad q_0\sdot q_0 \,g\sdot g > 0\,,
\end{equation}
and either
\begin{equation}
 \left( g\sdot q_0 \right)^2 < q_0\sdot q_0 \,g\sdot g\,,
\end{equation}
in which case there is only one real root, or
\begin{equation}
 \left( g\sdot q_0 \right)^2 \ge q_0\sdot q_0 \,g\sdot g\quad\mbox{and}\quad g\sdot q_0\, g_0 > 0\,,
\end{equation}
in which case all roots are negative. Further constraints on the parameters come from the requirement
\begin{equation}
 \hat f^{-1}H^{-1}-\hat f^2\omega_z{}^2 > 0\,,
\end{equation}
that also implies $H>0$.

The near horizon geometries of these black holes are themselves supersymmetric solutions and are included in the class of solutions we presented. They can be obtained
from equations (\ref{eq:c1_f}), (\ref{eq:c1_psi}) and (\ref{eq:c1_omz}) by taking the limit (\ref{eq:scaling_limit}) and choosing $q_{10}=0$. They are analogous to the three
near horizon geometries obtained in \cite{Gutowski:2004ez} for pure supergravity, in particular one can easily see from (\ref{eq:null_coord_metric}) that 
dimensional reduction along $v$ gives the geometries AdS$_2\times S^2$, AdS$_2\times \mathbb{H}^2$ or 
AdS$_2\times \mathbb{E}^2$, and that the horizon geometry
is given by a homogeneous Riemannian metric on the group manifolds SU(2) (in which case the metric is that of a squashed $S^3$), SL($2,\mathbb{R}$) or $Nil$
respectively for $k=1$, $-1$ or $0$. The entropy for the compact $k=1$ case was computed in \cite{Gutowski:2004yv}.

\subsubsection{Conserved charges}

For $k=1$ the class of solutions we presented is asymptotically globally AdS$_5$ according to the definition given by Ashtekar and Das in 
\cite{Ashtekar:1999jx}.\footnote{See \cite{Chimento:2016mmd} for a discussion of the asymptotics of a similar class of solutions in pure gauged supergravity.} It is then
possible to use the prescription in the same paper to compute the AD mass and angular momenta.

The mass is the conserved charge associated with the timelike Killing vector field
\begin{equation}
 V=\frac{\partial}{\partial t}+\frac{2}{\ell}\frac{\partial}{\partial z}\,.
\end{equation}
This is the correct vector rather than the one associated with supersymmetry, since in coordinates adapted to $V$ the metric of AdS$_5$, and in particular the metric on the 
conformal boundary, is written in static form. The value of the mass is
\begin{equation}
\begin{split}
 \mathcal{M}&=\frac{g_0 \ell^2}{2\sqrt{3}}g\sdot q_0+\frac{1}{8 g_0\ell}\frac{g_Ic_0{}^I}{(c_1{}^0)^2} q_0\sdot q_0\\
 &\\
   &+ \frac{3}{32\ell}\left(q_0\sdot q_0-\frac{32\ell}{3}c_1{}^0 \right)\left\{2g_0+\frac{g_Ic_0{}^I}{(c_1{}^0)^2}\left[\sqrt{3} g\sdot q_0
   +\frac{1}{\ell^3}\left(q_0\sdot q_0-\frac{32\ell}{3}c_1{}^0 \right)\right]  \right\}\,.
\end{split}
\end{equation}
Before computing the angular momenta, we perform the coordinate change
\begin{equation}
\label{eq:spher_coord}
 z=\psi+\varphi+\frac{2}{\ell}t\,,\qquad x^1=\tan\tfrac{\theta}{2}\cos\varphi\,,\qquad x^3=\tan\tfrac{\theta}{2}\sin\varphi\,,
\end{equation}
so that 
\begin{equation}
 dz+\chi_{(1)}=d\psi+\cos\theta d\varphi+\frac{2}{\ell} dt\,,\qquad d\Omega_{(2,1)}^2=d\theta^2+\sin^2\theta d\varphi^2\,.
\end{equation}
The angular momenta are the conserved charges associated with the Killing vectors $\partial_\varphi$ and $\partial_\psi$. They are
\begin{gather}
 J_\varphi=0\,,\\
 \nonumber\\
\label{eq:angular_mom_psi}
 J_\psi=\frac{1}{64}\left(q_0\sdot q_0-\frac{32\ell}{3}c_1{}^0 \right)\!\left[\frac{3}{\ell^3}\frac{g_Ic_0{}^I}{(c_1{}^0)^2}\left(q_0\sdot q_0-\frac{32\ell}{3}c_1{}^0 \right)
 +2\left(2g_0+\sqrt{3}\frac{g_Ic_0{}^I}{(c_1{}^0)^2}g\sdot q_0  \right)   \right]\,.
\end{gather}

The electric charges, defined by
\begin{equation}
 \mathcal{Q}_I=\frac{1}{8\pi G} \int_{S^3_\infty}a_{IJ}\ast F^J
\end{equation}
are
\begin{align}
\label{eq:Q0}
 \mathcal{Q}_0&= \frac{1}{128}\left[g_0\left(\!q_0\sdot q_0-\frac{32\ell}{3}c_1{}^0 \right)\left(2g_0-\sqrt{3}\frac{g_Ic_0{}^I}{(c_1{}^0)^2}g\sdot q_0  \right)
 -4\frac{g_Ic_0{}^I}{(c_1{}^0)^2}q_0\sdot q_0\right]\,,\\
 \nonumber\\
\label{eq:Qx}
 \mathcal{Q}_x&= -\frac{1}{128}\left\{\!\left(\!q_0\sdot q_0-\frac{32\ell}{3}c_1{}^0 \right)\!
 \left[2g_0g_x-\sqrt{3}\frac{g_Ic_0{}^I}{(c_1{}^0)^2}\left(g_x g\sdot q_0-g\sdot g q_{0x}\right) \right]\! +\frac{16}{\sqrt{3}} q_{0x}\right\}\,.
\end{align}

It is straightforward to verify that the following BPS condition is satisfied for all values of the parameters:
\begin{equation}
\label{eq:BPS}
 \mathcal{M}-\frac{2}{\ell}|J|=4 \ell |\tilde g^I \mathcal{Q}_I|
\end{equation}
where we have defined
\begin{equation}
 \tilde g^I\equiv \lim_{\varrho\to\infty}a^{IJ}g_J\quad\Rightarrow\quad \tilde g^0=\frac{1}{g_0 \ell^2}\,,\quad \tilde g^x=\frac{2}{\ell^2} \frac{g^x}{g\sdot g}\,.
\end{equation}
% the combination $\tilde g^I \mathcal{Q}_I$ appearing on the right hand side is the charge associated with the gauge field $g_I A^I$.

\subsubsection{Static solutions}

With the choice $c_1{}^0=\frac{3}{32 \ell}q_0\sdot q_0$ the functions $\Psi$ and $\omega_z$ can be expressed in a simple way in terms of $\hat f$,
\begin{align}
 \Psi&=\frac{4}{\ell^2}\varrho^3\hat f^{-3}+k\varrho^2\,,\\
 \nonumber\\
 \omega_z&=\frac{2}{\ell}\varrho\hat f^{-3}\,,
\end{align}
with $\hat f$ given by
\begin{equation}
 \hat f^{-3}=\frac{27}{2}\mathcal{H}_0 \mathcal{H}\sdot \mathcal{H}\,,
\end{equation}
where
\begin{equation}
 \mathcal{H}_I\equiv\frac{\ell}{3}g_I-\frac{\mathcal{Q}_I}{\varrho}
\end{equation}
and the $\mathcal{Q}_I$'s, that for $k=1$ are the electric charges (\ref{eq:Q0}) and (\ref{eq:Qx}), are
\begin{equation}
 \mathcal{Q}_0=-\frac{32 \ell^2}{9}\frac{g_Ic_0{}^I}{q_0\sdot q_0}\,, \qquad \mathcal{Q}_x=-\frac{q_{0x}}{8\sqrt{3}}\,.
\end{equation}
The gauge potentials and scalar fields can also be written in a simple way in terms of the functions $\mathcal{H}_I$,
\begin{equation}
 A^0=-\frac{dt}{3 \mathcal{H}_0}\qquad A^x=-\frac{2}{3}\frac{\mathcal{H}^x}{\mathcal{H}\sdot \mathcal{H}}dt\qquad \phi^x=\frac{\mathcal{H}_x}{\mathcal{H}_0}\,. 
\end{equation}

For $k=\pm 1$ it is possible to remove from the metric the cross term proportional to $dt(dz+\chi)$ by performing a simple shift in the $z$ coordinate, 
$z=\psi+\frac{2}{\ell k} t $, and rewrite the solutions as
\begin{equation}
 ds^2=\frac{\hat f^2}{k}\left( k+\frac{4}{l^2}\varrho \hat f^{-3} \right)dt^2-\frac{d\varrho^2}{\varrho \hat f \left( k+\frac{4}{l^2}\varrho \hat f^{-3} \right)}
 -\frac{\varrho}{\hat f}\left[k\left( d\psi+\chi_{(k)} \right)^2+ d\Omega_{(2,k)}^2 \right].
\end{equation}
Note that these coordinates are static for $k=1$ but not for $k=-1$, since in that case the time coordinate is actually $\psi$, while $t$ is spatial. 
However the metric can still be rewritten in static form making first the coordinate change
\begin{equation}
 \psi=\tilde\psi-\varphi\,,\qquad x^1=\tanh\tfrac{\theta}{2}\cos\varphi\,,\qquad x^3=\tanh\tfrac{\theta}{2}\sin\varphi\,,
\end{equation}
so that
\begin{equation}
 d\psi+\chi_{(-1)}=d\tilde\psi-\cosh\theta d\varphi\,,\qquad d\Omega_{(2,-1)}^2=d\theta^2+\sinh^2\theta d\varphi^2\,,
\end{equation}
followed by a second change,
\begin{equation}
 \tilde \psi=\alpha+\beta\,,\qquad \varphi = \alpha-\beta\,,\qquad \theta=2\vartheta\,,
\end{equation}
after which it takes the form
\begin{equation}
\begin{split}
 ds^2=-\hat f^2\left( -1+\frac{4}{l^2}\varrho \hat f^{-3} \right)dt^2&-\frac{d\varrho^2}{\varrho \hat f \left( -1+\frac{4}{l^2}\varrho \hat f^{-3} \right)}\\
 \\
 &-\frac{4\varrho}{\hat f}\left(-\cosh^2\vartheta d\beta^2+d\vartheta^2+\sinh^2\vartheta d\alpha^2 \right).
\end{split}
\end{equation}

For $k=1$ one can see that substituting the chosen value of $c_1{}^0$ in (\ref{eq:angular_mom_psi}) the angular momentum vanishes as expected. 
In this case the three-dimensional part of the metric contained in the square brackets is just the metric of a 3-sphere, with the coordinate change
\begin{equation}
 \psi=\tilde\psi+\varphi\,,\qquad x^1=\tan\tfrac{\theta}{2}\cos\varphi\,,\qquad x^3=\tan\tfrac{\theta}{2}\sin\varphi\,,
\end{equation}
one has
\begin{equation}
 \left( d\psi+\chi_{(1)} \right)^2+ d\Omega_{(2,1)}^2=4d\Omega_{S^3}^2=\left( d\tilde\psi+\cos\theta d\varphi \right)^2+d\theta^2+\sin^2\theta d\varphi^2\,.
\end{equation}
This solution was first found in \cite{Behrndt:1998ns}, and can be seen as a generalization in the presence of vector multiplets of the BPS limit of the 
Reissner-Nördstrom-AdS$_5$ black hole, to which it reduces in the pure supergravity case.

For $k=0$ it is not possible to eliminate the cross term in a simple way, and the metric is
\begin{equation}
 ds^2=\hat f^2 dt^2+\frac{4}{\ell}\frac{\varrho}{\hat f}dt\left( dz+\chi_{(0)} \right)-\frac{\ell^2}{4}\frac{\hat f^2 d\varrho^2}{\varrho^2}
 -\frac{\varrho}{\hat f}d\Omega_{(2,0)}^2\,.
\end{equation}
In the pure supergravity case this reduces to a metric without free parameters and having constant curvature scalars \cite{Chimento:2016mmd}. Here this is not true in 
general, and only happens if 
\begin{equation}
 \mathcal{H}\sdot\mathcal{H}=\frac{2}{(g_0)^3 \ell^3} (\mathcal{H}_0)^2\,,
\end{equation}
in which case the metric is the same as in the pure supergravity case, but it is still possible to have independent vector fields and non-trivial scalar fields.

\subsection{Case 2\label{subsec:case2}}

The solutions with
\begin{equation}
 q_{1x}=\frac{q_{10}}{g_0}\,g_x\,,\qquad c_1{}^0=g\sdot q_0=g\sdot c_1=q_0\sdot c_1=c_1\sdot c_1=g_Ic_0{}^I=0
\end{equation}
are almost identical to the black hole limit of the ones in Subsection \ref{subsec:case1}, given in equations (\ref{eq:bh_f}), (\ref{eq:bh_psi}) and (\ref{eq:bh_omz}), 
with the additional constraint $g\sdot q_0=0$. However there is an additional term in the 4-dimensional gauge potentials $\hat A^x$ proportional to the constants $c_1{}^x$, 
which were zero in the aforementioned limit. These constants are not completely arbitrary, being constrained by the relations $g\sdot c_1=q_0\sdot c_1=c_1\sdot c_1=0$.

After the rescaling (\ref{eq:q10_rescal}) the functions determining the metric are
\begin{align}
 \hat f^{-3}&=\left(1+\frac{\sqrt{3}}{8 g_0 \ell}\frac{q_{00}}{\varrho}\right)
 \left(1 +  \frac{3}{64 g\sdot g \ell^2} \frac{q_0\sdot q_0}{\varrho^2} \right)\,,\\
 \nonumber\\
 \Psi&=\varrho^2\left(\frac{4}{\ell^2} \varrho+ k+\frac{\sqrt{3}}{4} q_{00} g\sdot g \right)\,,\\
 \nonumber\\
 \omega_z&=\frac{2}{\ell}\varrho
 +\frac{3}{64 \varrho}\left(\frac{8\ell}{\sqrt{3}}q_{00} g\sdot g\varrho+\frac12 g_0 q_0\sdot q_0\right)\,,
\end{align}
while the scalars are
\begin{equation}
 \phi^x=\frac{h_x}{h_0}=\frac{h_x/\hat f}{h_0/\hat f}=\frac{8 g_x \ell \varrho+\sqrt{3} q_{0x}}{8 g_0\ell\varrho+\sqrt{3} q_{00}}\,,
\end{equation}
and the gauge potentials are of the form (\ref{eq:sol_full_gauge_pot}), with
\begin{align}
 \hat A^0&=\left(   g\sdot g\ell \varrho+\frac{1}{2}\frac{c_1{}^0}{\varrho}\right)\left( dz+\chi_{(k)} \right)\,,\\
 \nonumber\\
 \hat A^x&=\left[2 g_0 g^x\ell\varrho+\frac{\sqrt{3}}{4}\left( q_{00}g^x+ g_0q_0{}^x\right)
 + \frac{1}{2}\frac{c_1{}^x}{\varrho}\right]\left( dz+\chi_{(k)} \right)\,,
\end{align}
and
\begin{equation}
 h^0 \hat f = \frac{8 \varrho}{\sqrt{3} (8 g_0\,\ell \varrho+\sqrt{3} q_{00})}\,,\qquad 
 h^x \hat f = \frac{16\varrho}{\sqrt{3}} \frac{8 g^x \ell \varrho+\sqrt{3} q_0{}^x}{(8 g_y \ell \varrho+\sqrt{3} q_{0y})^2}\,.
\end{equation}

For $k=1$, the mass, angular momenta and electric charges are
\begin{align}
 \mathcal{M}&=\frac{q_{00}}{2\sqrt{3} g_0\ell}
   + \frac{3}{32\ell} q_0\sdot q_0\left(2g_0+\frac{4}{\sqrt{3}\ell^3}q_{00}  \right)\,,\\
   \nonumber\\
 J_\varphi&=0\,,\\
 \nonumber\\
 J_\psi&=\frac{q_0\sdot q_0}{16}\left(g_0+\frac{\sqrt{3}}{\ell^3}q_{00}\right)\,,\\
 \nonumber\\
 \mathcal{Q}_0&=\frac{1}{64}\left[(g_0)^2\ q_0\sdot q_0 -\frac{8}{\sqrt{3}}q_{00}\right]\,,\\
 \nonumber\\
 \mathcal{Q}_x&= -\frac{1}{64}\left[g_0 g_x q_0\sdot q_0+2 g\sdot g q_{00}q_{0x}+\frac{8}{\sqrt{3}}q_{0x}-\frac{32\ell}{3} g\sdot g\, c_{1x}\right]\,.
\end{align}
Keeping into account the constraints to which the constants $q_{0x}$ and $c_1{}^x$ are subject, it is easy to check that the relation (\ref{eq:BPS}) is satisfied.

%%%%%%%%%%%%%%%%%%%%%%%%%%%%%%%%%%%%%%%%%%%%%%%%%%%%%%%%%%%%%%%%%%%%%%
%%%%%%%%%%%%%%%%%%%%%%%%%%%%%%%%%%%%%%%%%%%%%%%%%%%%%%%%%%%%%%%%%%%%%%
%%%%%%%%%%%%%%%%%%%%%%%%%%%%%%%%%%%%%%%%%%%%%%%%%%%%%%%%%%%%%%%%%%%%%%
%%%%%%%%%%%%%%%%%%%%%%%%%%%%%%%%%%%%%%%%%%%%%%%%%%%%%%%%%%%%%%%%%%%%%%
\section{Conclusions}
\label{sec-conclusions}
%%%%%%%%%%%%%%%%%%%%%%%%%%%%%%%%%%%%%%%%%%%%%%%%%%%%%%%%%%%%%%%%%%%%%%
%%%%%%%%%%%%%%%%%%%%%%%%%%%%%%%%%%%%%%%%%%%%%%%%%%%%%%%%%%%%%%%%%%%%%%
%%%%%%%%%%%%%%%%%%%%%%%%%%%%%%%%%%%%%%%%%%%%%%%%%%%%%%%%%%%%%%%%%%%%%%
%%%%%%%%%%%%%%%%%%%%%%%%%%%%%%%%%%%%%%%%%%%%%%%%%%%%%%%%%%%%%%%%%%%%%%
In this paper we have adapted the equations that determine the timelike supersymmetric solutions of $\mathcal{N}=1$, $d=5$ Abelian gauged supergravity coupled to vector
multiplets to the assumption that the Kähler base space admits a holomorphic isometry. While the resulting system of equations is much more involved than in the pure
supergravity case, we were able, thanks in part to the experience gained in this latter case, to obtain several supersymmetric solutions.
Of these, the more interesting ones are three classes (for $k=0,\pm1$) of superficially asymptotically-AdS (globally asymptotically-AdS for $k=1$)
solutions, which are a direct generalization of the similar solutions found for pure supergravity in \cite{Chimento:2016mmd}, and which include various already known
solutions.

It is worth noting that the special geometric model \mbox{ST$[2,n_v+1]$} considered here admits as a special case the so-called U(1)$^3$ model, which is just the STU model
with equal gauging parameters $g_I$. This means that in this particular subcase our solutions can be oxidized to type-IIB supergravity as described in \cite{Cvetic:1999xp}.

The solutions constructed here only have one independent angular momentum, however there are in the literature examples of supersymmetric black holes with two independent
angular momenta in $\mathcal{N}=1$, $d=5$ Abelian gauged supergravity, both without and with vector multiplets \cite{Chong:2005hr, Kunduri:2006ek}. It would be interesting
to study whether less restrictive assumptions than those made in this paper could lead to solutions generalizing these black holes. Another possible extension of our work 
would be to consider more general gaugings, for instance a combination of the Abelian Fayet-Iliopoulos gauging considered here and non-Abelian gaugings of the 
scalar manifold isometries. Work along these lines is in progress \cite{future_work}.

%%%%%%%%%%%%%%%%%%%%%%%%%%%%%%%%%%%%%%%%%%%%%%%%%%%%%%%%%%%%%%%%%%%%%%
%%%%%%%%%%%%%%%%%%%%%%%%%%%%%%%%%%%%%%%%%%%%%%%%%%%%%%%%%%%%%%%%%%%%%%
%%%%%%%%%%%%%%%%%%%%%%%%%%%%%%%%%%%%%%%%%%%%%%%%%%%%%%%%%%%%%%%%%%%%%%
%%%%%%%%%%%%%%%%%%%%%%%%%%%%%%%%%%%%%%%%%%%%%%%%%%%%%%%%%%%%%%%%%%%%%%
\acknowledgments
%%%%%%%%%%%%%%%%%%%%%%%%%%%%%%%%%%%%%%%%%%%%%%%%%%%%%%%%%%%%%%%%%%%%%%
%%%%%%%%%%%%%%%%%%%%%%%%%%%%%%%%%%%%%%%%%%%%%%%%%%%%%%%%%%%%%%%%%%%%%%
%%%%%%%%%%%%%%%%%%%%%%%%%%%%%%%%%%%%%%%%%%%%%%%%%%%%%%%%%%%%%%%%%%%%%%
%%%%%%%%%%%%%%%%%%%%%%%%%%%%%%%%%%%%%%%%%%%%%%%%%%%%%%%%%%%%%%%%%%%%%%

The author would like to thank Tomás Ortín for his initial collaboration in this work,
useful comments and discussions. This work has been supported in part by the Spanish Ministry of Science 
and Education grants FPA2012-35043-C02-01 and FPA2015-66793-P 
(MINECO/FEDER, UE) and the Centro de Excelencia Severo Ochoa Program grant 
SEV-2012-0249.

\bibliography{papers}
\bibliographystyle{JHEP}

\end{document}